\documentclass{elsart}
\usepackage[dvips]{graphicx}
\usepackage{color}
\usepackage{epsfig,amsopn}

\def\tod{\mathop{{\to}}\limits}
\def\simd{\mathop{{\sim}}\limits}

\begin{document}
\begin{frontmatter}
\title
{Thermodynamics and dynamics of systems with long-range interactions}
\author{Freddy Bouchet$^{1,2}$},
\author{Shamik Gupta$^{3}$} and
\author{David Mukamel$^{3}$}

\address{1. INLN, CNRS, UNSA, 1361 route des lucioles, 06 560 Valbonne, France}
\address{2. LANL, CNLS, MS B258, PO Box 1663, Los Alamos, NM 87545, United States}
\address{3. Physics of Complex Systems, Weizmann Institute of Science, Rehovot 76100, Israel}

\date{\today}

\thanks[freddy]{E-mail: Freddy.Bouchet@inln.cnrs.fr}
\thanks[shamik]{E-mail: shamik.gupta@weizmann.ac.il}
\thanks[david]{E-mail: david.mukamel@weizmann.ac.il}
\begin{abstract}
We propose a lecture on simple aspects of the thermodynamic and dynamical properties of systems with long-range
pairwise interactions (LRI), which decay as $1/r^{d+\sigma}$ at large
distances $r$ in $d$ dimensions. Two broad classes of
such systems are discussed. (i) Systems with a slow decay of the
interactions, termed ``strong'' LRI, where the energy is
super-extensive. These systems are characterized by unusual
properties such as inequivalence of ensembles, negative specific
heat, slow decay of correlations, anomalous diffusion and ergodicity breaking. (ii) Systems with faster decay of the interaction potential, where the
energy is additive, thus resulting in less dramatic effects. These
interactions affect the thermodynamic behavior of systems near phase
transitions, where long-range correlations are naturally present.
Long-range correlations are often present in systems driven out of
equilibrium when the dynamics involves conserved quantities. Steady
state properties of driven systems with local dynamics
are considered within the framework outlined above.

\end{abstract}

\end{frontmatter}

\def\be{\begin{equation}}
\def\ee{\end{equation}}
\def\bea{\begin{eqnarray}}
\def\eea{\end{eqnarray}}
\def\bef{\begin{figure}}
\def\eef{\end{figure}}
\def\bml{\begin{mathletters}}
\def\eml{\end{mathletters}}
\def\l{\label}
\def\b{\bullet}
\def\no{\nonumber}
\def\fr{\frac}
\def\th{\theta}

\section{Introduction}
This paper provides a brief introduction to the thermodynamics and
dynamics of systems with long-range interactions (LRI). In these
systems, the interaction potential between the constituent particles
decays slowly with distance, typically as a power law $\sim
1/r^{d+\sigma}$ at large separation $r \gg 1$, where $d$ is the
spatial dimension. The interaction potential may be isotropic or
anisotropic (as in magnetic or electric dipolar systems). Long-range
interacting systems may be broadly classified into two groups: those
with $-d \le \sigma \le 0$, which are termed systems with ``strong''
LRI, and those with positive but not too large $\sigma$, which are
termed systems with ``weak'' LRI. Systems with strong LRI show
significant and pronounced dynamic and thermodynamic effects due to
the slow decay of the interaction potential. In contrast, in systems
with weak LRI, the potential decays relatively faster, resulting in
less pronounced effects. For a recent review on long-range
interacting systems, see \cite{Campa_2009}.

Long-range interacting systems are rather common in nature, for
example, self-gravitating systems ($\sigma=-2$)
\cite{Padmanabhan_1990}, non-neutral plasmas ($\sigma=-2$)
\cite{Nicholson_1991}, dipolar ferroelectrics and ferromagnets
(anisotropic interactions with $\sigma=0$) \cite{Landau_1960},
two-dimensional geophysical vortices ($\sigma=-2$)
\cite{Chavanis_houches_2002}, wave-particle interacting systems such
as a free-electron laser \cite{Barre_Dauxois_etal_2004_PRE_FEL}, and
many others.

Let us first consider systems with strong LRI. These systems are
generically non-additive, resulting in many unusual properties, both
thermal and dynamical, which are not exhibited by systems with weak
LRI or with short-range interactions. For example, the entropy may
turn out to be a non-concave function of energy, yielding negative
specific heat within the microcanonical ensemble
\cite{Antonov_1962}, \cite{LyndenBell:1968_MNRAS},
\cite{Lynden-Bell_1999}, \cite{Thirring_1970}, \cite{Hertel_1971}, \cite{Bouchet_Barre:2005_JSP}, \cite{Chavanis_2006IJMPB_Revue_Auto_Gravitant},
\cite{Posch_2006}. Since canonical specific heat is always positive,
it follows that the two ensembles need not be equivalent. More
generally, the inequivalence is manifested whenever a model exhibits
a first-order transition within the canonical ensemble
\cite{BarreMukamelRuffo:2001_PRL_BEC}, \cite{Barre_2002}.
Non-additivity may also result in breaking of ergodicity, where the
phase space is divided into domains. Local dynamics do not connect
configurations in different domains, leading to finite gaps in
macroscopic quantities such as the total magnetization in magnetic
systems \cite{Mukamel_2005}, \cite{Feldman_1998},
\cite{Borgonovi_2004}, \cite{Borgonovi_2006}, \cite{Hahn_2005},
\cite{Hahn_2006}, \cite{Bouchet_Dauxois_Mukamel_Ruffo_2007}.

Studies of relaxation processes in models with strong LRI have shown
that a thermodynamically unstable state relaxes to the stable
equilibrium state unusually slowly over a timescale which diverges
with the system size \cite{Mukamel_2005},
\cite{Antoni_Ruffo_1995PRE},
\cite{Latora_Rapisarda_Ruffo_1998_PhysRevLett_QSS?},
\cite{Latora_Rapisarda_Ruffo_1999_PhRvL_Superdiffusion},
\cite{Yamaguchi_2003_PRE},
\cite{Yamaguchi_Barre_Bouchet_DR:2004_PhysicaA}, \cite{Chavanis_houches_2002}, \cite{Spitzer_1991}, \cite{Dubin_ONeil_1999_RevModPhys_Revue_QSS}. This may be
contrasted with the relaxation process in systems with short-range
interactions. Diverging timescales in systems with strong LRI result
in long-lived quasistationary states. In the thermodynamic limit,
these states do not relax to the equilibrium state, so that the
system remains trapped in these states forever. These
quasistationary states and their slow relaxation have been
explained theoretically in the framework of kinetic theory
\cite{Spitzer_1991,Nicholson_1991,Dubin_ONeil_1999_RevModPhys_Revue_QSS,Chavanis_houches_2002,Yamaguchi_Barre_Bouchet_DR:2004_PhysicaA,Bouchet_Dauxois:2005_JOP}.
Recent progress in the kinetic theory of systems with long-range
interactions \cite{Bouchet:2004_PRE_StochasticProcess,Bouchet_Dauxois:2005_JOP,Bouchet_Dauxois:2005_PRE}
has also uncovered algebraic relaxation and explained anomalous
diffusion in and out of equilibrium.

It is worthwhile to point out that non-additivity may occur even in
finite systems with short-range interactions in which surface and
bulk energies are comparable.  Negative specific heat in small
systems (e.g., clusters of atoms) has been discussed in a number of
studies \cite{RMLynden-Bell_1995}, \cite{RMLynden-Bell_1996},
\cite{Gross_2000}, \cite{Chomaz_2002}.

Systems with weak LRI, for which $\sigma > 0$, are additive. Unless
one is in the vicinity of a phase transition, their thermodynamic
properties are similar to those with short-range interactions, e.g.,
the specific heat is non-negative, and the various statistical
mechanical ensembles are equivalent. Near a phase transition,
long-range correlations build up. These correlations affect the
universality class of a system near a continuous phase transition,
resulting in critical exponents which depend on the interaction
parameter $\sigma$.  Moreover, for these systems, the upper critical
dimension $d_c(\sigma)$ above which the critical behavior becomes
mean-field-like depends on $\sigma$ and has a smaller value than in
systems with short-range interactions for which $d_c=4$, see Refs. \cite{Fisher_1972}, \cite{Fisher_1974}. A system with weak LRI may
exhibit phase transitions in one dimension at a finite temperature,
which are otherwise forbidden in a system with short-range
interactions \cite{Dyson_1969a}, \cite{Dyson_1969b}.

So far we have discussed systems in equilibrium. Long-range
correlations may also build up in driven systems which reach a
non-equilibrium steady state that violates detailed balance. Quite
generally, such steady states in systems with conserving dynamics
exhibit long-range correlations, even with local dynamics. One thus
expects peculiarities in behavior of equilibrium systems with
long-range interactions to also show up in steady states of
non-equilibrium systems with conserving local interactions. An
example of such a non-equilibrium system with long-range
correlations is the so-called $ABC$ model. In this model, three
species of particles, $A$, $B$ and $C$, move on a ring with local
dynamical rules. At long times, the system reaches a nonequilibrium
steady state in which the three species are spatially separated. The
dynamics of this model lead to effective long-range
interactions in the steady state \cite{Evans_1998a},
\cite{Evans_1998b}.

The paper is laid out as follows. In Section \ref{strong}, we
discuss the thermodynamics and dynamics of systems with strong
long-range interactions. This is followed by a discussion on upper
critical dimension for systems with weak LRI in Section \ref{weak}.
The $ABC$ model, exhibiting long-range correlations under
out-of-equilibrium conditions is discussed in Section
\ref{sec:NESS-Local}. The paper ends with conclusions.

\section{Strong long-range interactions}
\l{strong}
\subsection{Thermodynamics}
\l{thermodynamics}
\bef
\begin{center}
\includegraphics[scale=0.4]{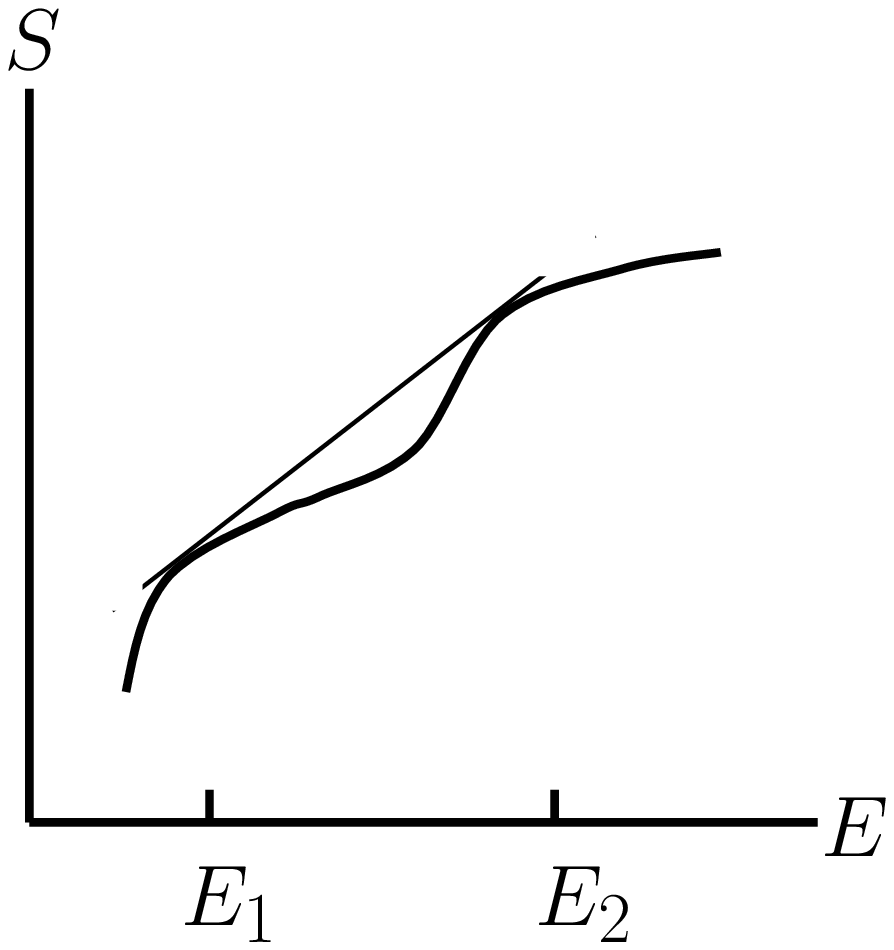}
\caption{Entropy as a non-concave function of energy. For
short-range systems, due to additivity, the physically realizable
curve in the interval $E_1 < E <E_2$ is given by the common tangent
line, resulting in an overall concave curve. In systems with
long-range interactions, the non-concave curve may be actually
realizable, giving rise to negative microcanonical specific heat.}
\l{Fig1}
\end{center}
\eef Here we briefly discuss some general thermodynamic properties
of systems with strong LRI. These systems are non-extensive and
non-additive. For example, the energy of a particle interacting with
a homogeneous distribution of particles in a volume $V$ scales as
$V^{-\sigma/d}$, so that the total energy scales superlinearly with
the volume as $V^{1-\sigma/d}$, making it non-extensive, and hence,
non-additive.

The most immediate consequence of non-additivity is that, unlike
short-range systems, the entropy $S$ is not necessarily a concave
function of energy. This may be understood by referring to Fig.
\ref{Fig1}. The equilibrium state at a given energy within a
microcanonical ensemble is obtained by maximizing the entropy at
that energy. A short-range interacting system is unstable in the
energy interval $E_1 < E < E_2$, since it can gain in entropy by
phase separating into two subsystems with energies $E_1$ and $E_2$,
keeping the total energy fixed. The energy and entropy densities are
then given by the weighted average of the corresponding densities of
the two coexisting subsystems. As a result, the physically
realizable entropy curve in the unstable region is obtained by the
common tangent line, resulting in an overall concave curve. However,
in systems with strong LRI, due to non-additivity, the energy
density of two coexisting subsystems is not given by the weighted
average of the energy density of the two subsystems. Therefore, the
non-concave curve of Fig. \ref{Fig1} could, in principle, represent
a physically realizable stable system, with no occurrence of phase
separation. This results in a microcanonical negative specific heat
in the interval $E_1 < E < E_2$ \cite{Antonov_1962},
\cite{LyndenBell:1968_MNRAS}, \cite{Lynden-Bell_1999},
\cite{Thirring_1970}, \cite{Hertel_1971}, \cite{Posch_2006}. Since
the specific heat within the canonical ensemble is always positive,
being given by the fluctuations about the   mean of the system
energy, this leads to inequivalence of ensembles, which is
particularly manifested whenever a first-order transition with
coexistence of two phases is found within the canonical ensemble
\cite{BarreMukamelRuffo:2001_PRL_BEC}, \cite{Barre_2002}.

Another feature related to non-additivity is that of a discontinuity
in temperature at a first-order phase transition within a
microcanonical ensemble, say, from a paramagnetic to a magnetically
ordered phase. This may be understood by referring to Fig.
\ref{Fig2}(a) which shows the entropy $S(M,E)$ as a function of the
magnetization $M$ at an energy $E$ close to the transition. It
exhibits three local maxima, one at $M=0$ and two other degenerate
maxima at $M=\pm M_0$. As the energy varies, the heights of the
peaks change. The paramagnetic phase occurs at energies such that
$S(0,E)>S(\pm M_0,E)$, while the magnetically ordered phase occurs
at energies where the inequality is reversed. The temperature in the
two phases are given by $1/T=\partial S(0,E)/\partial E$ and
$1/T=\partial S(\pm M_0,E)/\partial E$, respectively. At the
transition point, when one has $S(0,E)=S(\pm M_0,E)$, these two
derivatives are generically not equal, resulting in a temperature
discontinuity. This shows up in the entropy vs. energy curve, see
Fig. \ref{Fig2}(b).

The non-additive property also manifests itself in dynamical
features through breaking of ergodicity, the reason for which may be
traced back to the fact that in systems with strong LRI, the domain
in the phase space of extensive thermodynamic variables may be
non-convex. As a result, gaps may exist in phase space between two
points corresponding to the same energy, so that local
energy-conserving dynamics cannot take the system from one point to
the other, leading to breaking of ergodicity.

\bef
\begin{center}
\includegraphics[scale=0.35]{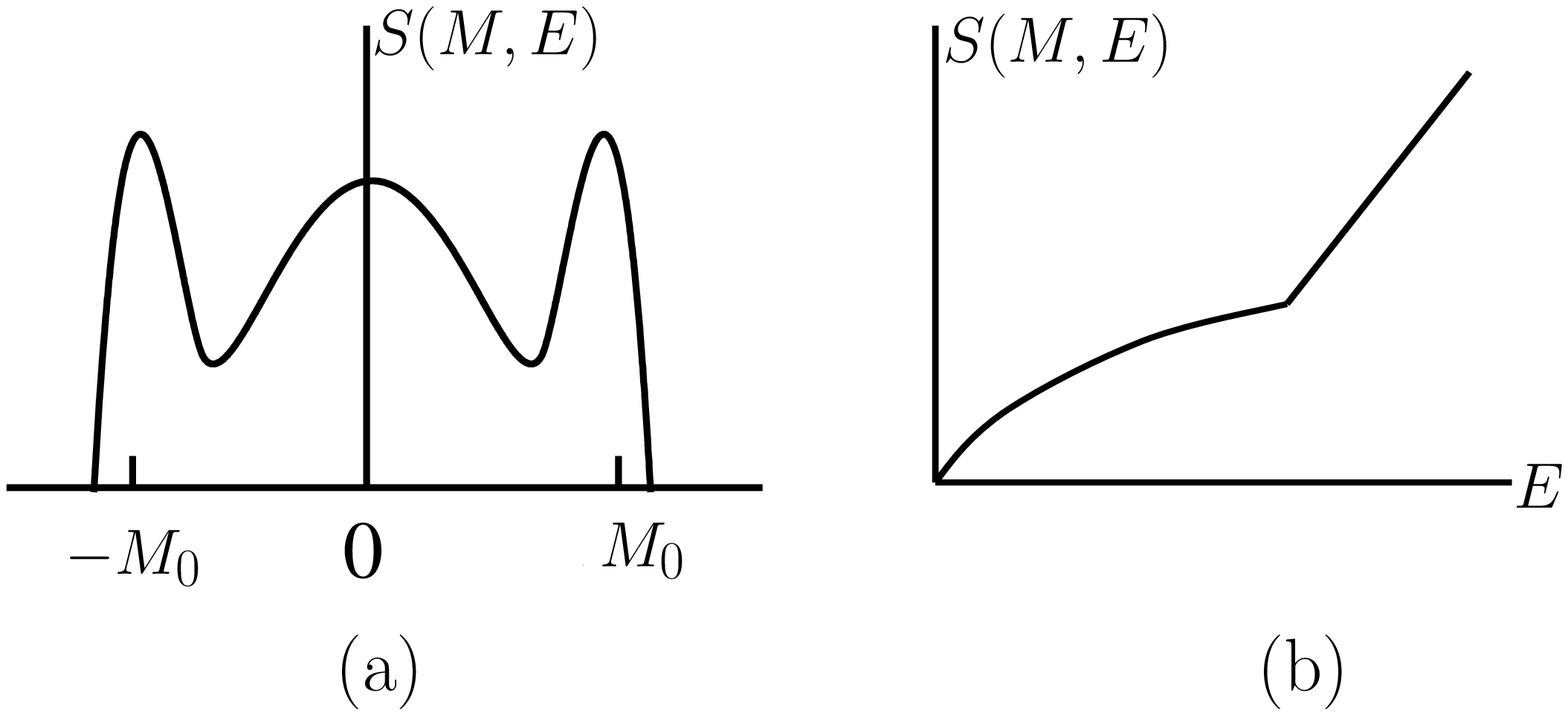}
\caption{(a) Entropy vs. magnetization close to a first-order
transition in a magnetic system with long-range interactions. (b)
Entropy vs. energy, showing a slope and hence, a temperature
discontinuity at a first-order transition point.} \l{Fig2}
\end{center}
\eef The entropy $S$ of a system, given by the number of ways of
distributing $N$ particles with total energy $E$ in a given volume
$V$, typically scales linearly with the volume for both short- and
long-range interacting systems. The energy, on the other hand,
scales superlinearly with the volume in systems with LRI. As a
result, in these systems in the thermodynamic limit, the dominant
contribution to the free energy $F=E-TS$ at any finite temperature
$T$ is due to the energy, resulting in a trivial thermodynamics with
the ground state always representing the equilibrium state. However,
there are examples of finite-sized real systems with long-range
interactions (e.g., self-gravitating systems such as globular
clusters, \cite{Chavanis_houches_2002}) where the temperature could be
sufficiently high to make the entropic term $TS$ compete with the
energy $E$, resulting in a non-trivial thermodynamics. To study this
limit in theory, it is convenient to rescale the energy by the
factor $V^{\sigma/d}$ (alternatively, rescale the temperature by
$V^{-\sigma/d}$) so that the two terms in $F$ become comparable.
This was first suggested by Kac \cite{Kac_1963}. Although such
rescaling makes the system extensive, it remains non-additive,
leading to unusual thermodynamic properties, as mentioned above.

To illustrate some of these unusual thermodynamic behavior in
systems with strong LRI, it is instructive to analyze phase diagrams
of representative models. A class of models amenable to exact
analysis comprises those where the long-range part of the
interaction is of mean-field type. These models have been applied to
study dipolar ferromagnets \cite{Campa_2007}. An example in this
class is the Ising model with both long- and short-range
interactions. The model considers Ising-spins $S_i=\pm 1$ on a
one-dimensional lattice of $N$ sites with periodic boundary
conditions. The Hamiltonian is given by \be
H=-\frac{K}{2}{\sum_{i=1}^N}\left(S_iS_{i+1}-1\right)-\frac{J}{2N}
\left({\sum_{i=1}^N}S_i\right)^2. \l{hamex} \ee Here the first term
represents a nearest-neighbor coupling which could be either
ferromagnetic $(K>0)$ or antiferromagnetic $(K<0)$. On the other
hand, the second term, corresponding to a long-range, mean-field
type interaction, is ferromagnetic, $J>0$.

\bef
\begin{center}
\includegraphics[scale=0.25]{Fig3.eps}
\caption{The $(K,T)$ phase diagram for the Hamiltonian in Eq.
(\ref{hamex}) within the canonical and the microcanonical ensembles.
Here $J=1$. In the canonical ensemble, the large $K$ transition is
continuous (bold solid line) down to the canonical tricritical point
CTP where it turns first order (dashed line). In the microcanonical
ensemble, the continuous transition coincides with the canonical one
at large $K$ (bold line). It persists at lower $K$ (dotted line)
down to the microcanonical tricritical point MTP where it becomes
first order, with a branching of the transition line (solid lines).
The shaded region between these two lines is not accessible. The figure is taken from \cite{Mukamel_2005}.}
\l{Fig3}
\end{center}
\eef The canonical phase diagram of this model was analyzed in
\cite{Nagle_1970}, \cite{Bonner_1971}, \cite{Kardar_1983}, while the
microcanonical phase diagram was obtained in \cite{Mukamel_2005}.
Figure {\ref{Fig3}} shows the phase diagram in the two ensembles in
the $(K,T)$ plane with $J=1$ and antiferromagnetic $K$. Here $T$ is
the temperature. We see that the microcanonical and the canonical
critical lines coincide up to the canonical tricritical point CTP.
The microcanonical line extends beyond this point into the region
where, within the canonical ensemble, the model is magnetically
ordered. In this region, the microcanonical specific heat is
negative. At $K=K_{MTP}$, the microcanonical transition turns first
order, with a branching of the transition line and a temperature
discontinuity. The shaded region in the phase diagram represents an
inaccessible domain resulting from the discontinuity in temperature.

On quite general grounds, one expects the above features of the
phase diagram to be valid for any system in which a continuous
transition line becomes first order at a tricritical point, for
example, in the phase diagrams of the spin-$1$ Blume-Emery-Griffiths
model \cite{BarreMukamelRuffo:2001_PRL_BEC,Barre_2002}, in an $XY$ model
with two- and four-spin mean-field-like ferromagnetic interaction
terms \cite{Buyl_2005}, and in an $XY$ model with long- and
short-range, mean-field type interactions
\cite{Campa_Giansanti_Mukamel_Ruffo_2006_PhyA}. A classification of
possible types of inequivalent canonical and microcanonical phase
diagrams in systems with long-range interactions is given in
\cite{Bouchet_Barre:2005_JSP}.

\subsection{Dynamics of Hamiltonian systems: Kinetic theories \label{sec:Kinetic}} We now turn to the dynamics of
systems with strong long-range interactions. The dynamics of
discrete spin systems with long-range interactions will be
considered in Section \ref{sec:Microcanonical_Monte-Carlo}. Here we
consider continuous Hamiltonian systems with long-range
interactions. For simplicity, we limit our discussion to the
following dynamical equations of motion for a system of $N$
particles, given by

\bea
\dot{x}_{i} & = & p_{i}, \nonumber \\
\label{eq:edo} \\
\dot{p}_{i} & = & -\frac{1}{N}\sum_{j\neq i}\frac{dW(x_i-x_j)}{dx_i}
\nonumber. \eea Here $x_i$ and $p_i$ are, respectively, the
coordinate and the momentum of the $i$-th particle and $W(x)$ is the
interparticle potential. For simplicity, we first discuss the case
where the potential $W$ is of infinite range, i.e.,
every particle interacts with every other (mean-field interaction).
The case $W\left(x\right)\sim 1/x^{d+\sigma}$, where $d$ is the
spatial dimension, is very similar, as long as $\sigma <1$. Here the
variable $x$ is a spatial variable, similar to the variable $r$ is
the previous section. In some cases, for instance, in the HMF model
discussed later (see Eq. (\ref{eq:HMF})), it could also be
interpreted as an internal degree of freedom. Note that, in
accordance with the prescription of Kac (Section
\ref{thermodynamics}), the potential in the Hamiltonian dynamics,
Eq. (\ref{eq:edo}), is scaled by the factor $1/N$. This scaling
factor arises from a change of timescale, and is the natural choice
here, as it implies that each particle experiences a force of $O(1)$
in the limit $N \rightarrow \infty$.

In the limit of large $N$, the dynamical evolution given by Eq. (\ref{eq:edo})
is well approximated by kinetic theories. On a relatively short
timescale (that diverges with $N$), the evolution is described by the
Vlasov equation. On a much longer timescale, the relaxation towards
equilibrium is governed by the Lenard-Balescu-type dynamics (or, its
approximation by the Landau equation). These equations have been
applied to self-gravitating stars, plasmas in the weak-coupling
limit, and point vortex models in two-dimensional turbulence
\cite{Spitzer_1991,Nicholson_1991,Chavanis_houches_2002,Bouchet_Dauxois:2005_PRE,Dubin_ONeil_1988_PhysRevLett_Kinetic_Point_Vortex,Dubin_2003_Phys_Plasma_Collisional_Diffusion_Point_Vortex}.

The aim of the following subsections is to briefly present these
classical kinetic equations and some recent results related to them,
including predictions of quasistationary states, anomalous diffusion
and algebraic relaxation. Our presentation makes a systematic
parallel to the well-known case of the Boltzmann equation for
short-range systems and also stresses numerous analogies between the
two cases.

\subsubsection{Comparison of kinetic theories of long-range and
short-range interacting systems} \l{sub:Kinetic_Parallel} The
kinetic theory of systems with short-range interactions in the
dilute gas limit involves the Boltzmann equation, a cornerstone of
classical statistical mechanics (see, for example,
\cite{Lifshitz_Pitaevskii_1981_Physical_Kinetics} for a physical
approach, or \cite{Spohn_1991} for a precise mathematical
discussion). Microscopically, particles travel at a typical velocity
$\bar{v}$ and collide with each other after traveling a typical
distance $l$, called the mean free path. Let $\sigma$ be the
diffusion cross-section for these collisions. One has
$\sigma=\pi a^{2}$, where the parameter $a$ is of the order of the particle
radius. The mean free path is defined as $l=1/\left(\pi
a^{2}n\right)$, where $n$ is the typical particle density. The
Boltzmann equation applies when the ratio $\Gamma=a/l$ is small (the
Boltzmann-Grad limit \cite{Spohn_1991}).

In the limit $\Gamma \rightarrow 0$, any two colliding particles can
be considered as independent (uncorrelated) as they come from very
distant areas. This is the basis of the Boltzmann hypothesis of
molecular chaos (Stosszahl Ansatz). It explains why the evolution
of the phase space distribution function
$f(\mathbf{x},\mathbf{p},t)$ may be described by an autonomous
equation, the Boltzmann equation, given by \be \frac{\partial
f}{\partial t}+\frac{\mathbf{p}}{m}\cdot\frac{\partial
f}{\partial\mathbf{x}}=\frac{\bar{v}}{l}C\left(f\right).
\l{eq:Boltzmann} \ee Here $m$ is the mass of the particles, while
$C\left(f\right)$ represents collisional interactions between
particles.

In what follows, we explain that, for systems with long-range
interactions in the limit of large $N$, any two particles become
statistically independent. This may seem paradoxical as, in this
case, the force on every particle is the result of its interaction
with all the other particles. The equivalent of the Stosszahl Ansatz
(the fact that two particles are independent to leading order in
$1/N$) here is then due to the law of large numbers: the force on
each particle being the result of a large number of contributions
from its interaction with all the other particles, the exact value
of each contribution is of little importance and correlation between
the motion of two particles is small. We explain this in more detail
in Sections \ref{sub:Vlasov_QSS} and \ref{sub:Lenard-Balescu}.

\begin{center}
\begin{table}[htbp]
\centering \begin{tabular}{|l|c|c|}
\cline{2-3}
\multicolumn{1}{l||}{} & Short-ranged dilute gases  & Long-range systems\tabularnewline
\hline
Small parameter  & $a/l=1/\left(\pi a^{2}n\right)$ & $1/N$\tabularnewline
\hline
Equations:  &  & \tabularnewline
Initial evolution  & Collisionless Boltzmann & Vlasov equation\tabularnewline
& equation  & \tabularnewline
Late relaxation & Boltzmann equation  & Lenard-Balescu equation \tabularnewline
towards equilibrium & & \tabularnewline
\hline
Vanishing correlations  & Yes  & Yes\tabularnewline
Boltzmann entropy  & Yes  & Yes\tabularnewline
Stosszahl Ansatz  & Yes  & Yes\tabularnewline
\hline
Steady states of  & Local Poisson distribution   & Quasistationary states\tabularnewline
 the initial evolution & or local thermal equilibrium & \tabularnewline
\hline
Relaxation timescale  & $\propto l/\bar{v}$ or larger  & $\propto N$ or larger\tabularnewline
\hline
Long-range & Yes  & Yes\tabularnewline
temporal correlations & Yes  & Yes\tabularnewline
and algebraic decays  &  & \tabularnewline
\hline
Anomalous diffusion  & Dimension dependent & Yes\tabularnewline
\hline
\end{tabular}
\caption{The Boltzmann equation on the one hand, and the Vlasov and
the Lenard-Balescu equations on the other hand are obtained in two
opposite limits: for the former, in the Grad limit for dilute gases
(rare collisions), while, for the latter, in the limit where each
particle interacts with a macroscopic number of others. The
structure of the kinetic theory in both cases, however, share many
analogies, as shown in the table.} \l{tab:Boltzmann}
\end{table}
\par\end{center}

The analogy between the kinetic theory of dilute gases and that of
systems with long-range interactions extend further. This is
summarized in Table \ref{tab:Boltzmann}. The Boltzmann equation has
a Lyapunov functional: the entropy, given by $-\int dxdp\, f\log f$,
can be proven to increase in time ($H$-theorem). According to the
classical argument, the entropy, $-\int dxdp\, f\log f$, is given by
the number of microstates corresponding to the phase space
distribution $f$, so long as two particles can be considered
independent (this is the case in the Boltzmann-Grad limit).
Moreover, it is a general property that, for a system in which the
evolution of the macroscopic phase space density $f$ is described by
an autonomous equation, the number of microscopic states
corresponding to $f$ has to increase in time
\cite{Goldstein_Lebowitz_2004PhyD..193...53G}. These two arguments
thus explain why the $H$-theorem must hold for the Boltzmann
equation. Since, for systems with long-range interactions, two
particles can also be considered independent (see the previous
paragraph), an $H$-theorem with the same entropy, given by $-\int dxdp\, f\log
f$, must also hold in this case. The long-time evolution of systems
with long-range interaction is governed by the Lenard-Balescu
equation (Section \ref{sub:Lenard-Balescu}), for which the entropy
increase can actually be checked directly.

For the Boltzmann equation, there is an initial dynamical stage,
independent of collisions, which is governed by the free transport
only (Eq. (\ref{eq:Boltzmann}), with the right hand side set to
zero), and leads to local Poisson statistics \cite{Spohn_1991}.
Similarly, evolution of long-range interacting systems for short
times leads to a state where two-point correlation functions are
negligible, as explained in Section \ref{sub:Vlasov_QSS}. In
short-range systems, when gradients of intensive parameters
(density, temperature, etc.) are small, one achieves, for dilute
gases, local thermodynamic equilibrium which is believed to hold in
the limit of long times \cite{Spohn_1991}. Similarly, systems with
long-range interactions, on times of order one, converge towards
``quasistationary states'' (Section \ref{sub:Vlasov_QSS}), which
then evolve very slowly towards global statistical equilibrium
(Section \ref{sub:Lenard-Balescu}). Quasistationary states for
long-range interacting systems are thus the analogue of local
thermodynamic equilibrium of the Boltzmann equation for short-range
systems.

Starting with Einstein's paper on Brownian motion, a very important
class of works tries to relate macroscopic diffusion properties to
microscopic correlations functions (Kubo-type formulae). An
important result of classical kinetic theory is the long-time
algebraic behavior of the correlation functions and Kubo integrands
\cite{Pomeau_Resibois_1975PhRepport}, and the related anomalous
diffusion \cite{Pomeau_Resibois_1975PhRepport}. This leads to
long-range temporal correlations of some statistical properties. As
has been recently discovered, similar behavior occurs also in
systems with long-range interactions
\cite{Bouchet:2004_PRE_StochasticProcess,Bouchet_Dauxois:2005_JOP,Bouchet_Dauxois:2005_PRE}.
We explain this in Section \ref{sub:Anomalous_diffusion}.

\subsubsection{Vlasov dynamics and quasistationary states\label{sub:Vlasov_QSS}}

We now derive heuristically the Vlasov equation from the Hamiltonian
dynamics, Eq. (\ref{eq:edo}). A particle with coordinate $x$ feels a
potential
$V_{\mathrm{discrete}}\left(x\right)=\frac{1}{N}\sum_{i}W\left(x-x_{i}\right)$.
It is natural to consider the following continuum approximation to
this potential: \be V(x,t)=\int dy dp W(x-y)f(y,p,t). \l{eq:phi} \ee
The time evolution for the one-particle phase space
distribution function $f(x,p,t)$ follows the Vlasov equation, given by
\be \frac{\partial f}{\partial t}+p\frac{\partial f}{\partial
x}-\frac{\partial V}{\partial x}\frac{\partial f}{\partial p}=0.
\l{eq:vlasov} \ee If $\left\{x_{i}\right\} $ were $N$ independent
random variables distributed according to the distribution $f$, Eq.
(\ref{eq:phi}) would then follow from the law of large numbers, and
would be a good approximation to $V_{\mathrm{discrete}}$ up to
corrections of order $1/\sqrt{N}$. Replacing the true discrete
potential by $V$ thus amounts to neglecting correlations between
particles (the equivalent of the Stosszahl Ansatz) and finite-$N$
effects. The potential $V_{\mathrm{discrete}}$ being replaced by an
average one, namely, $V$, may be seen as a mean-field approximation
to the dynamics.

That this approximation is valid in the limit $N\rightarrow\infty$
may be understood more precisely from a physical point of view in
two different ways: by either writing the
Bogoliubov-Born-Green-Kirkwood-Yvon (BBGKY) hierarchy, closing the
hierarchy by considering a systematic expansion in powers of $1/N$, and keeping terms to leading order, or, by following the
Klimontovich approach (see Section \ref{sub:Lenard-Balescu}). The
validity of the Vlasov equation has also been established with
mathematical rigor for smooth $W$
\cite{Braun_Hepp_CommMathPhys_1977} (see also \cite{Spohn_1991}, and
a more recent work, \cite{Hauray_Jabin_ARMA_2007}, for some classes
of singular potential). These exact results show that the Vlasov
equation is a good approximation to the particle dynamics, at least
for times much smaller than $\log N$. Recent results showed that
this $\log N$ is actually optimal, in the sense that there actually exist sets of initial conditions 
exhibiting divergence on times of order $\log N$
\cite{Jain_Bouchet_Mukamel_2007_JStatMech} (see an analogous $\log
N$ timescale arising in Monte-Carlo dynamics for discrete spin
systems considered in Section \ref{sec:Microcanonical_Monte-Carlo}).

However, the ``coincidence time'' between the Vlasov dynamics and
the Hamiltonian dynamics is generically much longer than $\log N$, meaning that most of the initial conditions have a ``coincidence time'' much longer than $\log N$.
As will be discussed below, generic initial conditions converge
towards a stable stationary state of the Vlasov equation on a timescale of order one, and then stay trapped close to this state for
times algebraic in $N$ (see
\cite{Yamaguchi_Barre_Bouchet_DR:2004_PhysicaA} for a numerical
observation and \cite{Caglioti_Rousset_2007_JStatPhys_QSS} for a
mathematical investigation of the phenomenon).

As can be easily verified, the Vlasov equation, Eq.
(\ref{eq:vlasov}), inherits the conservation laws of the Hamiltonian
dynamics, for instance, the energy \be H[f]=\int dx dp
\left[f\frac{p^{2}}{2}+\frac{f V\left[f\right]}{2}\right],
\l{eq:Energie_Continue} \ee and the linear or the angular momentum
when the system has the corresponding translational or rotational
symmetry, respectively. The functionals, \be C_{s}[f]=\int dx dp
~s\left(f(x,p,t)\right), \l{eq:casimirs} \ee sometimes called
Casimirs, are also invariant, \emph{for any function} $s$.

Let us consider a dynamical system $\mathcal{F}$:
$\dot{x}=F\left(x\right)$, with a conserved quantity
$G\left(x\right)$ ($\dot{G}$$\left(x\right)=0$). Any extremum
$x_{0}$ of $G$ represents an equilibrium of $\mathcal{F}$:
$F\left(x_{0}\right)=0$ and if, in addition, the second variations
of $G$ are either positive definite or negative definite, then this
equilibrium is stable \cite{Holm_etal_PhysRep_1985}. This general
result seems natural if one considers the example of energy and
angular momentum extrema encountered in classical mechanics. Then,
as a consequence of the infinite number of conserved quantities,
Eqs. (\ref{eq:Energie_Continue}-{\ref{eq:casimirs}), there exists an
infinite number of equilibria $f_{0}$ for the Vlasov dynamics, a
large number of them being stable
\cite{Yamaguchi_Barre_Bouchet_DR:2004_PhysicaA}. In any dynamical
system, fixed points play a major role. In the case of the Vlasov
equation, they moreover turn out to be attractive, as illustrated by
Landau damping \cite{Nicholson_1991}. Following these simple
remarks, the following dynamical scenario was proposed
\cite{Yamaguchi_Barre_Bouchet_DR:2004_PhysicaA}:
\begin{itemize}
\item Starting from some initial condition, the $N$-particle system approximately follows the Vlasov dynamics, and evolves on a
timescale of order $1$.
\item It then approaches a stable stationary state of the Vlasov equation. Subsequently, the Vlasov evolution stops (``quasistationary states'').
\item Because of discreteness effects, the system evolves on a timescale of order $N^{\alpha}$ for some $\alpha$, and slowly approaches the statistical equilibrium, moving along a series of stable stationary states of the Vlasov equation (see Section \ref{sub:Lenard-Balescu}).
\end{itemize}
We note that a similar scenario was also observed in the plasma \cite{Nicholson_1991,Dubin_ONeil_1999_RevModPhys_Revue_QSS}, astrophysical \cite{Spitzer_1991} or point vortex \cite{Chavanis_houches_2002} context. As a concrete example, let us consider the case of the Hamiltonian
mean-field (HMF) model, which involves classical $XY$ spins with
mean-field interactions \cite{Antoni_Ruffo_1995PRE}. Here the
Hamiltonian is given by \be
H=\sum_{i=1}^{N}\frac{p_{i}^{2}}{2}+\frac{1}{2N}\sum_{i,j=1}^{N}\left[1-\cos\left(\theta_{i}-\theta_{j}\right)\right],
\l{eq:HMF} \ee and the magnetization is given by
$M=\left|\frac{1}{N}\sum_{i=1}^{N}\exp\left(i\theta_{i}\right)\right|$. Note that, for the HMF model, the variables $\theta_i$'s play the role of
the variables $x_i$'s in the discussion following Eq.
(\ref{eq:edo}). For the HMF model, the above mentioned dynamical
scenario is actually observed for initial states which are
homogeneous in angles $\theta_i$ and uniform in momenta $p_i$
(water-bag initial condition)
\cite{Latora_Rapisarda_Ruffo_1998_PhysRevLett_QSS?,Yamaguchi_Barre_Bouchet_DR:2004_PhysicaA}
(see Fig. \ref{fig:Magnetization_HMF}).

\begin{figure}[htbp]
\centering \includegraphics[height=0.2\textheight,keepaspectratio]{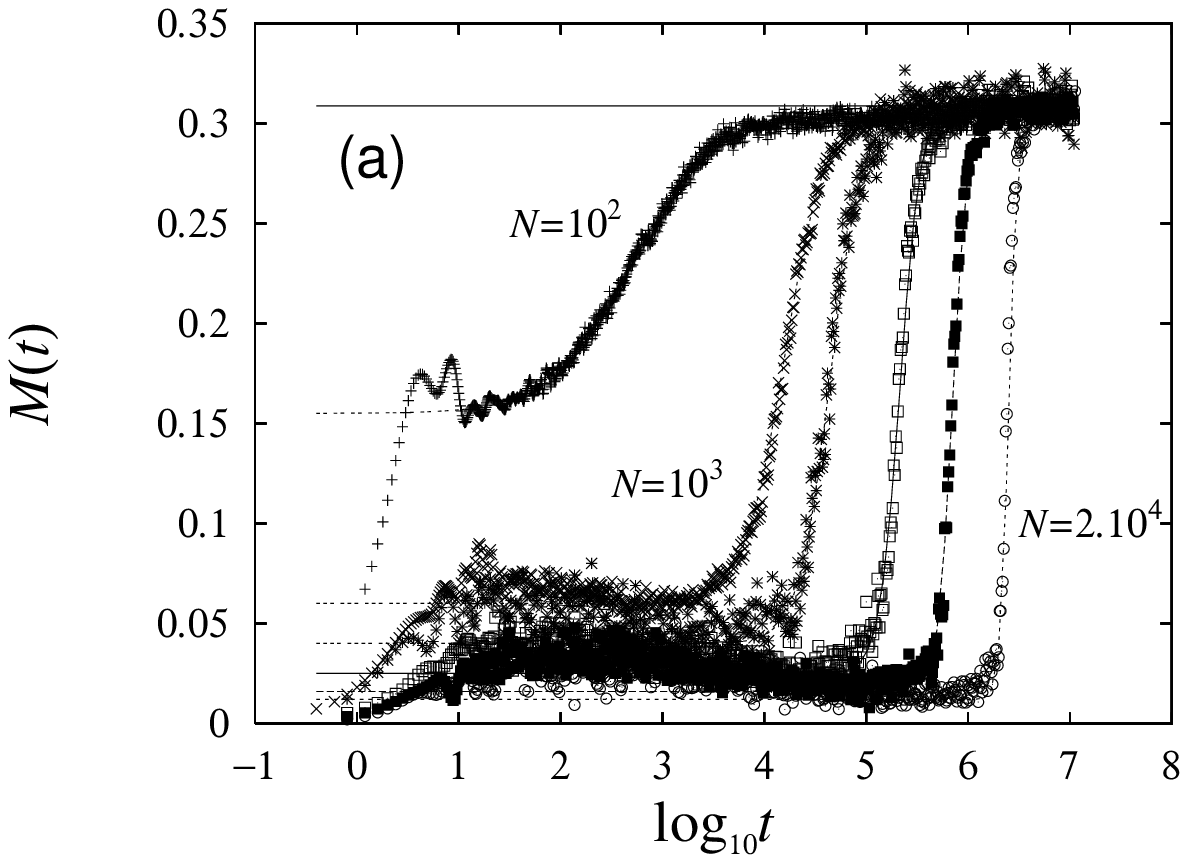}
\includegraphics[height=0.2\textheight,keepaspectratio]{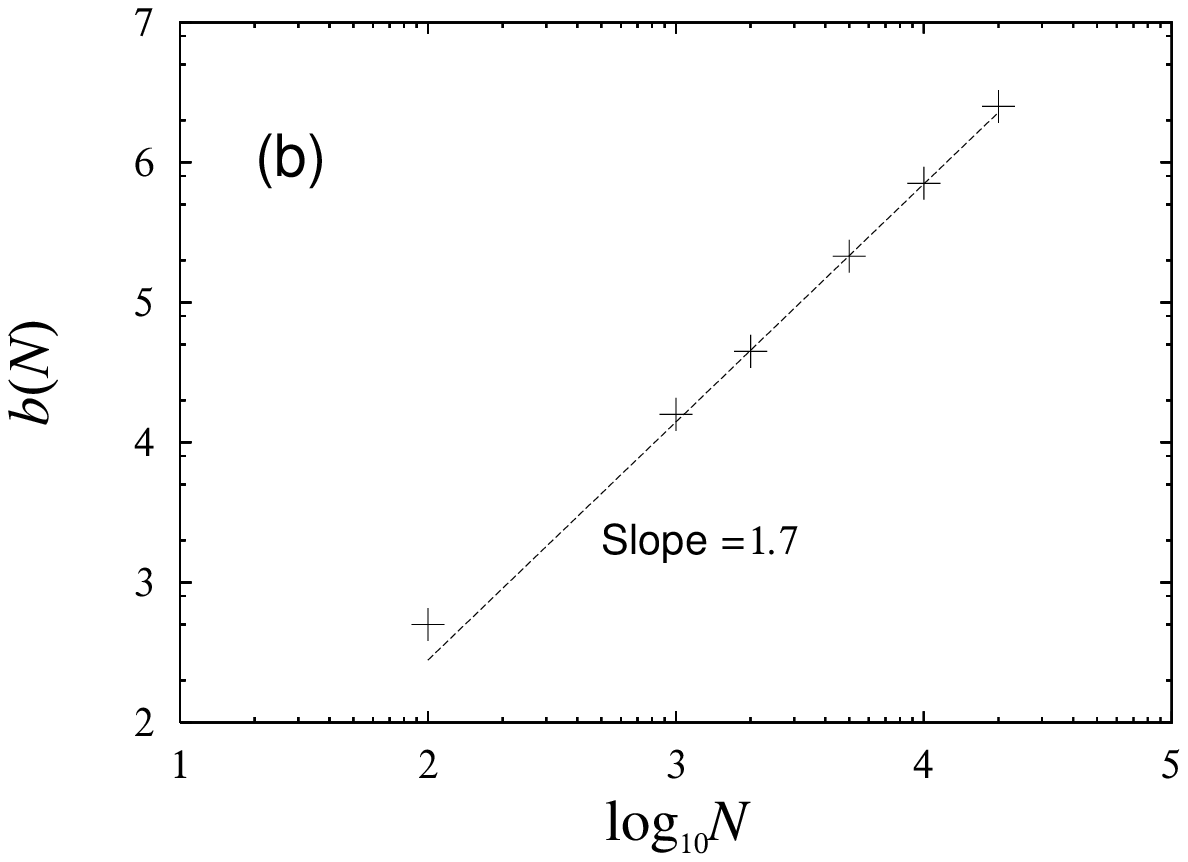}
\caption{Panel (a): Magnetization $M(t)$ of the HMF model, Eq. (\ref{eq:HMF}), for different particle numbers: from left to right, $N=10^{2}$,
$10^{3}$, $2.10^{3}$, $5.10^{3}$, $10^{4}$ and $2.10^{4}$. The initial state is homogeneous in angles $\theta_i$ and uniform in momenta $p_i$. The horizontal line at the top represents the statistical equilibrium value of $M$. Panel (b) shows the logarithm of the relaxation timescale $b(N)$ as a function of $\ln N$, where the dashed line represents the law $10^{b(N)}\sim N^{1.7}$. The figure is taken from \cite{Yamaguchi_Barre_Bouchet_DR:2004_PhysicaA}.
\label{fig:Magnetization_HMF}}
\end{figure}

In this scenario, the $N$-particle system gets trapped for long
times in out-of-equilibrium states close to stable stationary states
of the Vlasov equation; these are called quasistationary states
(QSS) in the literature. Before turning to a discussion of these
states in the next paragraph, let us note that there is however no
reason for this scenario to be the only possibility. For instance,
the Vlasov dynamics could converge towards stable periodic solutions
of the Vlasov equation \cite{Firpo_etal_PRE_2001}.

We have explained that any Vlasov-stable stationary solution is a
quasistationary state. Then, because inhomogeneous Vlasov-stationary
states do exist, one should not expect quasistationary states to be
homogeneous. This is illustrated in the case of several
generalizations of the HMF model in
\cite{Jain_Bouchet_Mukamel_2007_JStatMech}.

The issue of the robustness of QSS when the Hamiltonian is perturbed
by short-range interactions
\cite{Campa_Giansanti_Mukamel_Ruffo_2006_PhyA}, or, when the system
is coupled to an external bath
\cite{BaldovinOrlandini:2006_PRL_ThermalBath} has also been
addressed, and it was found that, while the power law behavior
survives at least on some timescale, the exponent may not be
universal. A possible statistical mechanical explanation of these
QSS would be the ``violent relaxation'' theory of Lynden-Bell
\cite{LyndenBell:1968_MNRAS} and its generalizations. We refer to
\cite{Chavanis_PhysicaA_2006,Barre_Dauxois_etal_2004_PRE_FEL} and
references therein for discussions on the interests and limitations
of this approach.

\subsubsection{Order parameter fluctuations and the Lenard-Balescu
equation\label{sub:Lenard-Balescu}}

In the previous subsection, we explained that, to leading order in
$1/\sqrt{N}$, the dynamical evolution is described by the Vlasov equation. We
now treat the $1/\sqrt{N}$ fluctuations of the order parameter, and
the resulting correlations and corrections to the Vlasov equation.
We assume that the initial condition is close to a QSS, and that
this property holds in time as the system evolves (this is the
equivalent of the propagation of local equilibrium for the Boltzman
equation \cite{Spohn_1991}).

In order to keep this discussion simple, we treat the case of the
HMF model, Eq. (\ref{eq:HMF}). The case of a more general potential, Eq. (\ref{eq:edo}), can be treated following exactly the same procedure. We follow
\cite{Bouchet_Dauxois:2005_PRE}, and refer to
\cite{Lifshitz_Pitaevskii_1981_Physical_Kinetics} for a plasma
physics treatment, to
\cite{Dubin_ONeil_1988_PhysRevLett_Kinetic_Point_Vortex,Dubin_2003_Phys_Plasma_Collisional_Diffusion_Point_Vortex,Chavanis_houches_2002}
for the case of point vortices, and to
\cite{Binney_Tremaine_1987_Galactic_Dynamics} for self-gravitating
stars.

A way to perform these computations would be an asymptotic expansion
of the BBGKY hierarchy, where $1/\sqrt{N}$ is the small parameter
(see, for instance, \cite{Nicholson_1991}). The $1/\sqrt{N}$
fluctuations would then be obtained by explicitly solving the
dynamical equations for the two-point correlation function while
truncating the BBGKY hierarchy by assuming a Gaussian closure for
the three-point correlation function. Our presentation, giving the
same results, rather follows the Klimontovich approach
\cite{Nicholson_1991,Lifshitz_Pitaevskii_1981_Physical_Kinetics}.

The state of the system of $N$ particles can be described by the {\em
discrete} single particle time-dependent density function
$f_{d}\left(\theta,p,t\right)$, defined as $f_{d}(\theta,p,t) \equiv
\frac{1}{N}\sum_{j=1}^{N}\delta\left(\theta-\theta_{j}\left(t\right)\right)\delta\left(p-p_{j}\left(t\right)\right)$,
where $\delta$ is the Dirac delta function, $(\theta,p)$ the Eulerian
coordinates of the phase space and $(\theta_{j},p_{j})$ the Lagrangian
coordinates of the particles. By taking the time derivative of
$f_{d}\left(\theta,p,t\right)$ and using Eq. (\ref{eq:edo}), one finds
that the dynamical evolution is described by the Klimontovich equation
\cite{Nicholson_1991}, given by
\be \frac{\partial f_{d}}{\partial
t}+p\frac{\partial f_{d}}{\partial \theta}-\frac{dV}{d\theta}\frac{\partial
f_{d}}{\partial p}=0, \l{equationfdiscrete} \ee with \be
V(\theta,t)\equiv-\int_{0}^{2\pi}d\theta'\int_{-\infty}^{\infty}dp \cos(\theta-\theta')
f_{d}(\theta',p,t). \l{eq:V_fd} \ee Equation (\ref{equationfdiscrete}) is
the same as the Vlasov equation, Eq. (\ref{eq:vlasov}) (with $x$ replaced by $\theta$). However,
whereas Eq. (\ref{equationfdiscrete}) describes the evolution of a
sum of Dirac distributions and is exact, the Vlasov equation
describes a smooth distribution $f$ understood as a local spatial
average (or a temporal average, depending on the interpretation).

When $N$ is large, it is natural to approximate the discrete density
$f_{d}$ by a continuous one, namely, $f\left(\theta,p,t\right)$. Considering an
ensemble of microscopic initial conditions close to the same initial
macroscopic state, one defines the statistical average $\langle
f_{d}\rangle=f_{0}(\theta,p)$, while fluctuations are of order $1/\sqrt{N}$. We will assume that $f_{0}$ is
any stable stationary solution of the Vlasov equation. The discrete
time-dependent density function can thus be written as
$f_{d}(\theta,p,t)=f_{0}(\theta,p)+\delta f(\theta,p,t)/\sqrt{N}$, where the
fluctuation $\delta f$ is of zero average. Similarly, we define the
average potential $\langle V\rangle$ and its corresponding
fluctuation $\delta V(\theta,t)$ so that $V(\theta,t)=\langle V\rangle+\delta
V(\theta,t)/{\sqrt{N}}$. Inserting both expressions in the Klimontovich
equation, Eq. (\ref{equationfdiscrete}), and taking the average, one
obtains
\begin{eqnarray}
\frac{\partial f_{0}}{\partial t}+p\frac{\partial f_{0}}{\partial \theta}-\frac{d\langle V\rangle}{d\theta}\frac{\partial f_{0}}{\partial p} & = & \frac{1}{N}\left\langle \frac{d\delta V}{d\theta}\frac{\partial\delta f}{\partial p}\right\rangle .
\l{equationpourfzero}
\end{eqnarray}
The above equation with the right hand side set to zero is the
Vlasov equation. The exact kinetic equation, Eq.
(\ref{equationpourfzero}), suggests that the quasistationary states
of Section \ref{sub:Vlasov_QSS} do not evolve on timescales much
smaller than $N$; this explains the extremely slow relaxation of the
system towards statistical equilibrium.

Let us now concentrate on stable homogeneous distributions
$f_{0}(p)$. Then, one has $\langle V\rangle=0$. Subtracting Eq.
(\ref{equationpourfzero}) from Eq. (\ref{equationfdiscrete}) and
using $f_{d}=f_{0}+\delta f/\sqrt{N}$, one gets
\begin{eqnarray}
\frac{\partial\delta f}{\partial t}+p\frac{\partial\delta f}{\partial \theta} & - & \frac{d\delta V}{d\theta}\frac{\partial f_{0}}{\partial p}=\frac{1}{\sqrt{N}}\biggl[\frac{d\delta V}{d\theta}\frac{\partial\delta f}{\partial p}-\left\langle \frac{d\delta V}{d\theta}\frac{\partial\delta f}{\partial p}\right\rangle \biggr].
\label{eq:Linearized_Vlasov}
\end{eqnarray}
For times much smaller than $\sqrt{N}$, we may drop the right hand
side encompassing quadratic terms in the fluctuations. The
fluctuating part $\delta f$ is then described by the left hand side
of Eq. (\ref{eq:Linearized_Vlasov}), which is the linearized Vlasov
equation.

The linearized Vlasov equation can be solved explicitly by
introducing the spatio-temporal Fourier-Laplace transform of $\delta
f$ and $\delta V$. This leads to \be \widetilde{\delta
V}(\omega,k)=-\frac{\pi\left(\delta_{k,1}+\delta_{k,-1}\right)}{\,\varepsilon(\omega,k)}\int_{-\infty}^{+\infty}\!\!
dp\ \frac{\widetilde{\delta f}(0,k,p)}{i(pk-\omega)},
\l{eq:solution_linearized_Vlasov} \ee where the dielectric
permittivity $\epsilon$ is given by \be
\epsilon(\omega,k)=1+\pi{k}\left(\delta_{k,1}+\delta_{k,-1}\right)\int_{-\infty}^{+\infty}\!\!
dp\ \frac{{\displaystyle \frac{\partial f_{0}}{\partial
p}}}{(pk-\omega)}. \ee Equation
(\ref{eq:solution_linearized_Vlasov}) describes exactly the
fluctuations to leading order. From it, we can compute any quantity
of interest, for instance, the potential autocorrelation or the
right hand side of Eq. (\ref{equationpourfzero}). We describe the
results without reproducing here the computational details (which
are long and tedious, see
\cite{Bouchet_Dauxois:2005_PRE,Lifshitz_Pitaevskii_1981_Physical_Kinetics}).

\paragraph{Potential autocorrelation}

For homogeneous states, by symmetry, one has
$\langle\widetilde{\delta V}(\omega_{1},k_{1})\widetilde{\delta
V}(\omega_{2},k_{2})\rangle=0$, except when $k_{1}=-k_{2}=\pm1$. For
$k=\pm 1$, one gets, after a transient exponential decay, the
general result \be \left\langle {\delta V}(t_{1},\pm1){\delta
V}(t_{2},\mp1)\right\rangle=\frac{\pi}{2}\int_{\mathcal{C}}d\omega\
e^{-i\omega(t_{1}-t_{2})}\frac{f_{0}(\omega)}{\left|\varepsilon(\omega,1)\right|^{2}}.
\l{correldeltaVfinal} \ee This is an exact result to leading order.

\paragraph{Lenard-Balescu equation}

In order to describe the slow evolution of the distribution $f_{0}$
due to finite-$N$ effects, we evaluate the right hand side of Eq.
(\ref{equationpourfzero}) to order $1/N$. This is, for systems with
long-range interactions, the analogue of the collision operator for
the Boltzmann equation for dilute systems with short-range
interactions. This collision operator is called the Lenard-Balescu
operator and it leads to the Lenard-Balescu equation, given by
\bea
\frac{\partial f_{0}(p,t)}{\partial t}&=&-\frac{1}{N}\frac{\partial}{\partial p} LB[f], \nonumber \\
LB[f]&=&\int dp'\frac{1}{\left|\epsilon(1,1)\right|} \left(f_{0}\left(p\right)\frac{\partial f_{0}}{\partial p}(p')-f_{0}\left(p'\right)\frac{\partial f_{0}}{\partial p}(p)\right)\delta\left(p-p'\right). \nonumber \\
\l{eq:Lenard_Balescu}
\eea
We have presented the computation of the Lenard-Balescu equation for the HMF model, where variables $\theta $ and $p$ are one dimensional. The generalization of this computation to the general potential as in Eq. (\ref{eq:edo}), and for variables $\mathbf{x} $ and $\mathbf{p}$ of dimensions larger than one leads to
\be
\frac{\partial f_{0}(\mathbf{p},t)}{\partial t}=-\frac{1}{N}\frac{\partial}{\partial\mathbf{p}} LB[f],
\ee
\bea
\hspace{-1cm} LB[f]=\int d\mathbf{k}d\mathbf{p}'\frac{\phi(k)}{\left|\epsilon(k,\mathbf{k}.\mathbf{p}')\right|} \mathbf{k}.\left(f_{0}\left(\mathbf{p}\right)\frac{\partial f_{0}}{\partial\mathbf{p}}(\mathbf{p}')-f_{0}\left(\mathbf{p}'\right)\frac{\partial f_{0}}{\partial\mathbf{p}}(\mathbf{p})\right)\delta\left(\mathbf{k}.\left(\mathbf{p}-\mathbf{p'}\right)\right). \nonumber \\
\l{eq:Lenard_Balescugen}
\eea
Here $\mathbf{k}$ is a wave vector,
$\phi(k)$ is the Fourier transform of the potential $V(\mathbf{x})$,
and $\left|\epsilon(k,\mathbf{k}.\mathbf{p}')\right|$ is the
dielectric permittivity. We note that the Lenard-Balescu operator is
a quadratic one, as is the collision operator $C(f)$ in the
Boltzmann equation, Eq. (\ref{eq:Boltzmann}). Moreover, this
operator involves a resonance condition through the Dirac
distribution
$\delta\left(\mathbf{k}.\left(\mathbf{p}-\mathbf{p'}\right)\right)$.

From Eq. (\ref{eq:Lenard_Balescu}), we expect a relaxation towards equilibrium of
any quasistationary state with a characteristic time of order $N$.
We note that, for plasma or self-gravitating systems, due to the
small distance divergence of the interaction potential, the
Lenard-Balescu operator diverges at small scales. This is
regularized by introducing a small scale cut-off. This leads to a
logarithmic correction to the relaxation time, which is then the
Chandrasekhar time for stellar systems, proportional to $N/\log N.$

One clearly finds from Eq. (\ref{eq:Lenard_Balescu}) that the
mechanism for the evolution of the distribution function is related
to two-particle resonances. An essential point is that the resonance condition
$p-p'=0$ cannot be fulfilled. This is because, with $p=p'$, the Lenard-Balescu operator, Eq. (\ref{eq:Lenard_Balescu}), which is odd in the variable $p$, would vanish.

 For physical systems for which $\mathbf{x}$ is a one-dimensional variable,
this proves that Vlasov-stable distribution functions do not evolve
on timescales smaller or equal to $N$. This is an important result:
\textit{generic out-of-equilibrium distributions, for
one-dimensional systems, evolve on timescales much larger than $N$} \cite{Kadomtsev_Pogutse_1970PhRvL}.
As noted in \cite{Bouchet_Dauxois:2005_PRE}, this explains why for the HMF model, relaxation does not occur on times scales of order $N$ (a $N^{1.7}$ scaling law was numerically observed in the HMF model
\cite{Yamaguchi_Barre_Bouchet_DR:2004_PhysicaA}, see Fig. \ref{fig:Magnetization_HMF}, page \pageref{fig:Magnetization_HMF}). A similar kinetic blocking due to the same type of lack of resonances may also occur in the case of the point vortex model \cite{Chavanis_2001PhRvE_64_PointsVortex}.

\subsubsection{The stochastic process of a single particle\label{single_particle}}

Let us now consider the relaxation properties of a test-particle,
labeled by $1$, surrounded by a background of $(N-1)$ particles with
a homogeneous distribution $f_0(p)$. We want to describe the
stochastic process of particle $1$. We will first prove that the
dynamics of this particle may be described by a Fokker-Planck equation.
For this, we generalize the computations of Section
\ref{sub:Lenard-Balescu}. As in Section \ref{sub:Lenard-Balescu}, for the sake of simplicity, we treat here the case of the HMF model, but extensions to the general case, Eq. (\ref{eq:edo}), is straightforward.

We first compute the diffusion $\left<\left(p_1(t)-p_1(0)\right)^2\right>$, where $p_1(0)$ and
$p_1(t)$ are the momentum of particle $1$ at initial time and at time $t$, respectively. Here the brackets denote averaging over the initial positions and
momenta of the remaining $N-1$ particles.  Taking into account the
knowledge of the position of particle $1$, the distribution $f_{d}$
(see Eq. (\ref{equationfdiscrete})) is
$f_{d}(\theta,p,t)=f_{0}(\theta,p)+\delta
f(\theta,p,t)/\sqrt{N}+\delta(\theta-\theta_1,p-p_1)/N$, where $\delta f$ is the
zero-average fluctuation of the density of the remaining $N-1$ particles. We define the average potential $\left< V\right>$ and its corresponding fluctuation $\delta V(\theta,t)$ so that
$V(\theta,t)=\left<V\right>+\delta V(\theta,t)/{\sqrt{N}}$. Then, from Eq. (\ref{eq:V_fd}), we obtain \be \delta
V(\theta,t)=-\int_{0}^{2\pi}d\theta'\int_{-\infty}^{+\infty}dp\,
\cos(\theta-\theta')\,\delta
f(\theta',p,t)-\frac{1}{\sqrt{N}}\cos\left( \theta-\theta_{1}\right).
\l{fluctpotentialbis} \ee Using the equations of motion, Eq. (\ref{eq:edo}), for the test particle and omitting from now on the
label $1$ for the sake of simplicity, one obtains \be
p(t)=p(0)-\frac{1}{\sqrt{N}}\int_{0}^{t}\!\! du \frac{d\delta
V}{d\theta}(u,\theta(u)). \l{pt} \ee Then \bea \left<\left(p(t)-p(0)\right)^{2}\right>
= \frac{1}{N}\int_{0}^{t}\int_{0}^{t}\!\! du du' \left<\frac{d\delta
V}{d\theta}(u,x(0))\frac{d\delta
V}{d\theta}(u',\theta(0))\right>+\mathrm{O}\left(\frac{1}{N}\right). \nonumber \\ \eea In deriving the above equation, we have replaced $\theta(u)$ by $\theta(0)$ in Eq. (\ref{pt}), which is
valid to leading order in $1/N$. From Eq. (\ref{fluctpotentialbis}), it
is clear that the average autocorrelation of the potential does not
depend on particle $1$ to leading order in $1/N$. Then, to leading
order, Eq. (\ref{correldeltaVfinal}) for the Laplace transform of the potential
autocorrelation can be used. We obtain \be
\left<\left(p(t)-p(0)\right)^{2}\right>  \simd_{t\to+\infty}
\!\!\frac{2t}{N}\, D(p), \l{differentmoments2} \ee where the
diffusion coefficient $D(p)$ is explicitly computed from Eq.
(\ref{correldeltaVfinal}). One gets
\be
D(p)=2\,\mbox{Re}\int_{0}^{+\infty}dt\ e^{ipt}\,\left\langle {\delta
V}(t,1){\delta V}(0,-1)\right\rangle =\
\frac{{\pi^{2}}f_{0}(p)}{\left|\varepsilon(p,1)\right|^{2}}.
\l{diffusioncoefficient} \ee The computation of
$\langle\left(p(t)-p(0)\right)\rangle$ is less straightforward as
then the corrections to the potential due to particle $1$ have to be
evaluated to next order. These computations are not conceptually
difficult (see
\cite{Bouchet:2004_PRE_StochasticProcess,Bouchet_Dauxois:2005_JOP,Bouchet_Dauxois:2005_PRE}),
but are too long to be presented here. We obtain \be
\left<\left(p(t)-p(0)\right)\right>  \simd_{t\to+\infty}
\!\!\frac{t}{N}\left(\frac{dD(p)}{dp}+\frac{1}{f_{0}}\frac{\partial
f_{0}}{\partial p}D(p)\right).\quad\null\l{differentmoments1} \ee

As the changes in the momentum $p$ are small (of order
$1/\sqrt{N}$), the description of the stochastic process in momentum $p$
by a Fokker-Planck equation is valid (see
\cite{Van_Kampen_1992_Book_StochasticProcesses}). The Fokker-Planck
equation is then characterized by the temporal behavior of the first
two moments, $\left<\left(p(t)-p(0)\right)^{n}\right>$; $n=1,2$
\cite{Van_Kampen_1992_Book_StochasticProcesses}. Rescaling the time
variable $\tau=t/N$, as suggested by Eqs. (\ref{differentmoments1})
and (\ref{differentmoments2}), the Fokker-Planck equation describing
the time evolution of the distribution of the test particle is \be
\frac{\partial f_{1}(\tau,p)}{\partial\tau}=\frac{\partial}{\partial
p}\left[D(p)\left(\frac{\partial f_{1}(\tau,p)}{\partial
p}-\frac{1}{f_{0}}\frac{\partial f_{0}}{\partial
p}f_{1}(\tau,p)\right)\right]. \l{fokkerplanckequation} \ee We
stress that this equation depends on the bath distribution $f_{0}$.
It is valid for both equilibrium baths (Gaussian $f_0$) and
out-of-equilibrium baths, provided that $f_{0}$ is a stable
stationary solution of the Vlasov equation. It is easily checked
that $f_1(p)=f_0(p)$ is the stationary solution to Eq.
(\ref{fokkerplanckequation}). Then, in the limit $\tau\to\infty$, the
test particle probability density function $f_{1}$ converges towards
the quasistationary distribution of the surrounding bath $f_{0}$.
This is consistent with the result that $f_{0}$ is stationary for
timescales of order $N$.

The Vlasov equation, the Lenard-Balescu equation and the
Fokker-Planck equation for a test particle are all classical
results. Recent results are the ones related to the understanding of
the importance of QSS and their extensive study in the context of the HMF model. 
The Fokker-Planck diffusion coefficient has also been tested numerically
\cite{Bouchet:2004_PRE_StochasticProcess}. In the next subsection, we explain
other recent results related to the very interesting and peculiar
properties of the Fokker-Planck equation, Eq.
(\ref{fokkerplanckequation}), and the associated algebraic temporal
correlation and anomalous diffusion.

\subsubsection{Long-range temporal correlations and anomalous diffusion\label{sub:Anomalous_diffusion}}

The quest for relations between observable macroscopic transport
properties and microscopic properties are at the core of the program
of equilibrium and out-of-equilibrium statistical mechanics.
Historically, this has played an important role, not only from a
practical point of view to have access to microscopic information
without observing them directly, but also from a conceptual point of
view. The Kubo-type formulae are an essential part of the theory,
relating microscopic correlation functions to diffusion
coefficients. In the 1970's, it came as a great surprise to discover
that the integrand of the Kubo formulae may diverge and lead to
anomalous diffusion and transport (see below). This led to a series
of very interesting papers reviewed in \cite{Pomeau_Resibois_1975PhRepport}.

We briefly recall that, when looking at the statistics of a spatial
variable $x$ as a function of time, when its moment of order $n$,
$\left\langle x^n(\tau)\right\rangle$, scales like $\tau^{n/2}$ at
long times, the associated transport is called {\em normal}.
However, {\em anomalous} transport
\cite{Bouchaud_Georges_1990_PhysRep,Castiglione_Mazzino_Muratore_Vulpiani_1999PhyD_Anomalous_Diffusion},
where moments do not scale as in the normal case, is also known in
some stochastic models, in continuous time random walks (Levy
walks), in kinetic theory \cite{Pomeau_Resibois_1975PhRepport} and
for systems with a lack of stationarity of the corresponding
stochastic process \cite{Bouchet_Cecconi_Vulpiani:2004_PRL}.

In this subsection, we present recent results
\cite{Bouchet_Dauxois:2005_PRE} that predicted the existence of
non-exponential relaxation, autocorrelation of the momentum $p$ with
algebraic decay at long times, and anomalous diffusion of the
spatial or angular variable $x$. These results thus show that,
similar to the case of the classical theory of systems with
short-range interactions \cite{Pomeau_Resibois_1975PhRepport},
anomalous transport exists also in the kinetic theory of systems with
long-range interactions. These results also clarify the highly
debated disagreement between different numerical simulations
reporting either anomalous
\cite{Latora_Rapisarda_Ruffo_1999_PhRvL_Superdiffusion} or normal
\cite{Yamaguchi_2003_PRE} diffusion, in particular, by delimiting
the time regime for which such anomalous behavior should occur.
These theoretical predictions have been numerically checked in
\cite{Yamaguchi_Bouchet_Dauxois_2007_JSMTE_Anomalous_Diffusion}.
Some recent results, extending this work, have also been reported for
the point vortex model \cite{Chavanis_Lemou_2005_PRE}. We note that
an alternative explanation, with which we disagree, both for the
existence of QSS and for anomalous diffusion has been proposed in
the context of Tsallis non-extensive statistical mechanics
\cite{Latora_Rapisarda_Tsallis_2001_PhRvE,Pluchino_Latora_Rapisarda_2004_PhyD}
(see \cite{Yamaguchi_Barre_Bouchet_DR:2004_PhysicaA} and
\cite{Bouchet_Dauxois:2005_PRE,Bouchet_Dauxois_DR:2006_EurPhysNews}
for further discussions).

Our results have been obtained by analyzing theoretically the
properties of the Fokker-Planck equation, Eq.
(\ref{fokkerplanckequation}), derived in Section
\ref{single_particle}. From a physical point of view, as particles
with large momenta $p$ move very fast in comparison to the typical
timescale of the fluctuations of the potential, they experience a
very weak diffusion and thus maintain their large momentum during a
very long time (one finds from Eq. (\ref{diffusioncoefficient}),
using $\left|\varepsilon(p,1)\right|^{2}\tod_{p\rightarrow\infty}1$,
that the diffusion coefficient decays as fast as the bath
distribution $f_{0}\left(p\right)$ for large times). Because of this
very weak diffusion for large $p$, the distribution of waiting time
for passing from a large value of $p$ to a typical value of $p$ is a
fat distribution. This explains the algebraic asymptotics for the
correlation function. From a mathematical point of view, these
behaviors are linked to the fact that the Fokker-Planck equation,
Eq. (\ref{fokkerplanckequation}), has a continuous spectrum down to
its ground state (without gap). This leads to a non-exponential
relaxation of different quantities and to long-range temporal
correlations
\cite{Bouchet_Dauxois:2005_JOP,Bouchet_Dauxois:2005_PRE}. These
results apply to the kinetic theory of any system for which the slow
variable (here the momentum) lives in an infinite space. By
explicitly deriving an asymptotic expansion of the eigenvalues and
eigenfunctions of the Fokker-Planck equation, the exponent for the
algebraic tail of the autocorrelation function of momentum has been
theoretically computed
\cite{Bouchet_Dauxois:2005_JOP,Bouchet_Dauxois:2005_PRE}. The
detailed analysis is a bit complex and tedious, thus, cannot be
reproduced here (a detailed presentation can be found in
\cite{Bouchet_Dauxois:2005_JOP}).

Let us present the results in the context of the HMF model, Eq.
(\ref{eq:HMF}), for which algebraic long-time behavior for momentum
autocorrelation has been first numerically observed in Refs.
\cite{Latora_Rapisarda_Tsallis_2001_PhRvE,Pluchino_Latora_Rapisarda_2004_PhyD}.
In its QSS, the theoretical law for the diffusion of angles
$\sigma_{\theta}^{2}(\tau)$ has also been derived in
\cite{Bouchet_Dauxois:2005_JOP,Bouchet_Dauxois:2005_PRE}. The
predictions for the diffusion properties are listed in Table
\ref{tab:Cp} and illustrated using numerical simulations of the HMF model in Fig. \ref{fig:diffusion_anomale}.
\begin{table}[htbp]
 \centering \begin{tabular}{|l|c|c|c|}
\hline
Tails of the bath   & $f_{0}(p)$  & $C_{p}(\tau)$  & $\sigma_{\theta}^{2}(\tau)$\tabularnewline
distribution function $f_0$. & $(|p|\to\infty)$  &  $(\tau\to\infty)$  &  $(\tau\to\infty)$ \tabularnewline
\hline
Power-law.  & $|p|^{-\nu}$  & $\tau^{-\alpha}$  & $\tau^{2-\alpha}$ \tabularnewline
\hline
Stretched exponential.  & $\quad\exp(-\beta|p|^{\delta})$\quad{}  & \quad{}$\frac{(\ln\tau)^{2/\delta}}{\tau}$\quad{}  & \quad{}$\tau(\ln\tau)^{2/\delta+1}$\quad{}\tabularnewline
\hline
\end{tabular}
\caption{Theoretical predictions of the autocorrelation function $C_{p}(\tau)$ of the momentum $p$ and of the standard deviation $\sigma_{\theta}^{2}(\tau)$ of the variable $\theta$ in the long-time regime, for different bath distributions $f_0(p)$. The results are valid for any bath distribution $f_0(p)$ which is strictly decreasing as $|p| \to \infty$, and depend on $f_0(p)$ only through its large $p$ asymptotic behavior (tail of the distribution). The prediction for $\alpha$ is $\alpha=(\nu-3)/(\nu+2)$. See Fig. \ref{fig:diffusion_anomale} for an illustration of these results using numerical simulations of the HMF model and Refs. \cite{Bouchet_Dauxois:2005_JOP,Bouchet_Dauxois:2005_PRE} for more details.}

\label{tab:Cp}
\end{table}

When the distribution $f_{0}(p)$ is changed within the HMF model, a
transition between weak anomalous diffusion (normal diffusion with
logarithmic corrections) and strong anomalous diffusion is predicted
(Table~\ref{tab:Cp}). We have numerically confirmed this theoretical
prediction
\cite{Yamaguchi_Bouchet_Dauxois_2007_JSMTE_Anomalous_Diffusion}. For
initial distributions with power-law or Gaussian tails, correlation
functions and diffusion are in good agreement with numerical
results. Diffusion is indeed {\em anomalous super-diffusion} in the
case of power-law tails, while {\em normal} when Gaussian. In the
latter case, the system is in equilibrium, but the diffusion
exponent shows a slow logarithmic convergence to unity due to a
logarithmic correction to the correlation function. The long
transient times before observing normal diffusion, even for Gaussian
distribution and in equilibrium, suggests that one should be very
careful to decide whether diffusion is anomalous or not from
numerical simulations.

\begin{figure}
\begin{centering}
\includegraphics[scale=0.6]{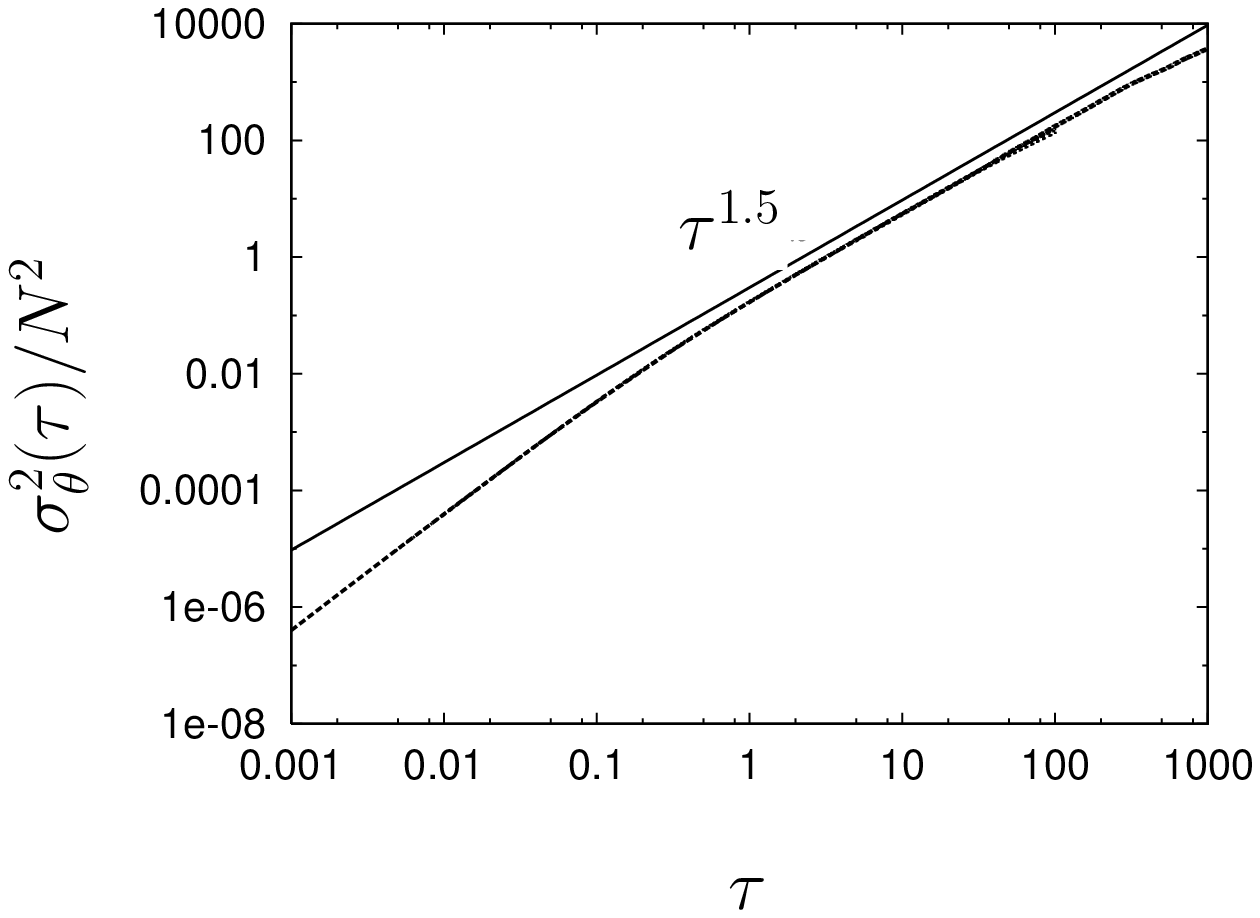}
\par\end{centering}
\caption{Diffusion ($\sigma_{\theta }^2(\tau)/N^2 \equiv \left< (\theta(t)-\theta(0))^{2} \right>/N^2$ as a function of time $\tau=t/N$) in the HMF model, for a quasistationary state. Points are from a $N$-body numerical simulation, the straight line is the analytic prediction by the kinetic theory. For long times, $\left<(\theta(t)-\theta(0))^{2}\right>\simd_{t\rightarrow\infty}t^{\nu}$ with $\nu\neq1$. A weak anomalous diffusion is also observed in equilibrium (see \cite{Bouchet_Dauxois:2005_PRE} for details).}
\l{fig:diffusion_anomale}
\end{figure}

\subsection{Dynamics of discrete spin systems \label{sec:Microcanonical_Monte-Carlo}}

In this section, we discuss the relaxation process from a
thermodynamically unstable state in long-range interacting systems with discrete degrees of freedom. These systems do not have intrinsic dynamics and one has to
resort to Monte Carlo (MC) dynamics within either a microcanonical
or a canonical ensemble. Here we briefly discuss the results for
the Ising model with long- and short-range interactions, defined by
the Hamiltonian in Eq. (\ref{hamex}) \cite{Mukamel_2005}.

Within a microcanonical ensemble, the dynamics followed in
\cite{Mukamel_2005} is based on the microcanonical MC algorithm of
Creutz \cite{Creutz_1983}. In this algorithm, an extra degree of
freedom, called the demon, with energy $E_D \geq 0$ samples
microstates of the system with energy $E-E_D$ by attempting random
single spin flips. At long times, to leading order in the system
size $N$, the distribution of $E_D$ attains the Boltzmann form,
$P(E_D) \sim \exp(-E_D/k_BT)$, where $T$ is the temperature of the
system with energy $E$. As long as the entropy of the system
increases with its energy, the temperature is positive and the
average energy of the demon is finite and small compared with the
system energy, with the latter scaling with $N$. The system energy
at any given time is $E-E_D$, with finite fluctuations.

In applying the above dynamics to models with long-range
interactions, one should note that, to next order in $N$,
$P(E_D)\sim \exp{(-E_D/T-E_D^2/{2C_VT^2})}$, where $C_V=O(N)$ is the
system's specific heat. In systems with short-range interactions,
the specific heat is non-negative so that the next-to-leading term
in the distribution is a stabilizing factor which is negligible for
large $N$. On the other hand, in systems with long-range
interactions, $C_V$ may be negative in some regions of the phase
diagram, and on the face of it, the next-to-leading term may
destabilize the distribution. However, the next-to-leading term is
small, of $O(1/N)$, so that as long as the entropy increases with
the energy, the next-to-leading term does not destabilize the
distribution.

The above dynamics has been applied to the model in Eq.
(\ref{hamex}). It was found that, starting with a zero magnetization
thermodynamically unstable state at energies where this state is a
local minimum of the entropy, the model relaxes to the equilibrium,
magnetically ordered state on a timescale which diverges with the
system size as $\ln N$ (see Fig. \ref{FigtaulnN}). \bef
\begin{center}
\includegraphics[scale=0.25]{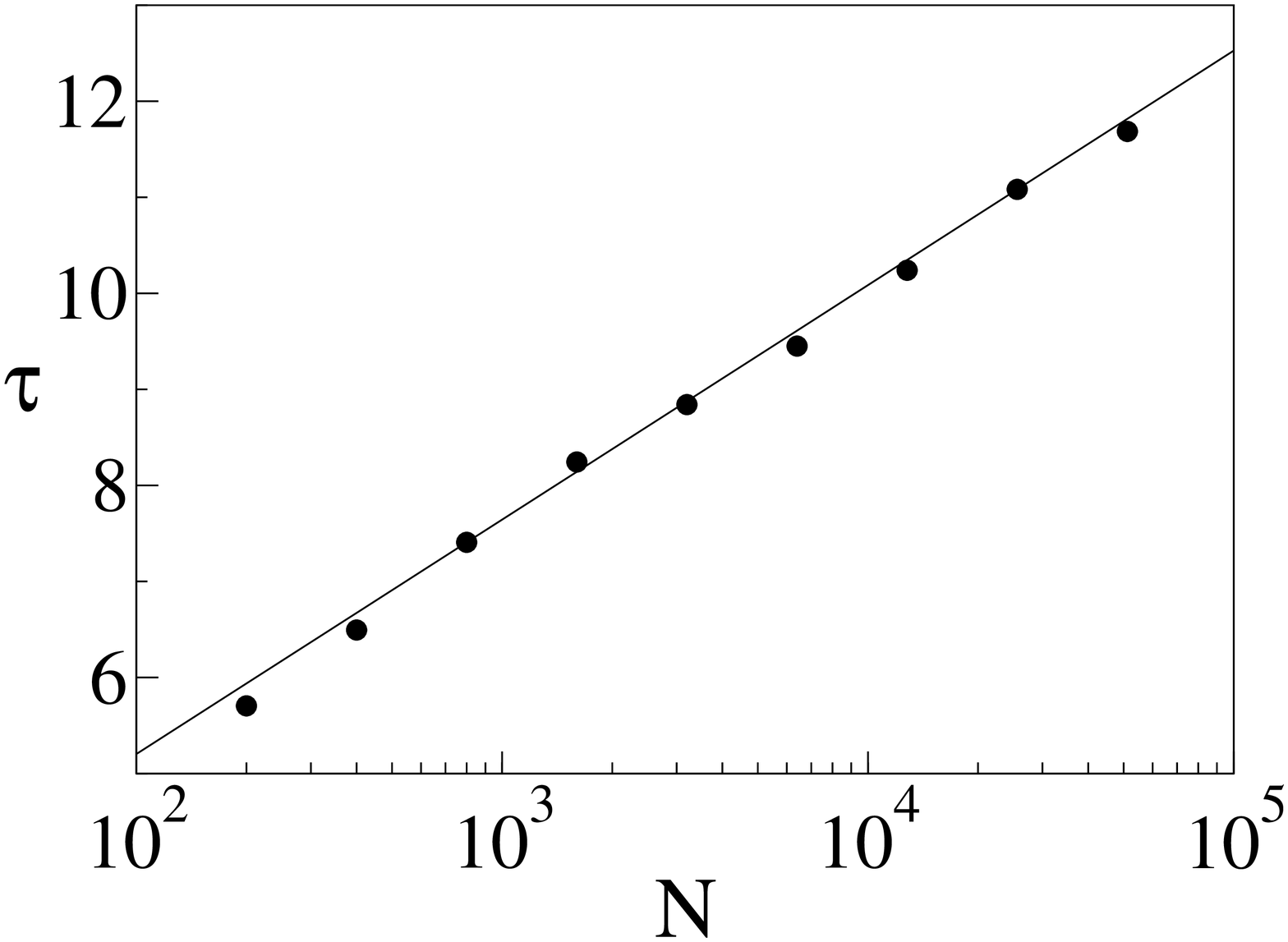}
\caption{Relaxation time of the $m=0$ state when this is a local
minimum of the entropy for the model in Eq. (\ref{hamex}). Here
$K=-0.25, J=1, \epsilon=-0.2$.} \l{FigtaulnN}
\end{center}
\eef

To get insight into the above result for the relaxation time, we
consider the Langevin equation corresponding to the dynamics, which
is given by \be \frac{\partial m}{\partial t}=\frac{\partial
s}{\partial m} +\xi (t); ~~~~~~~~ \langle \xi(t) \xi(t')\rangle = D
\delta(t-t'), \l{Langevin} \ee where $\xi(t)$ represents white
noise. The diffusion constant $D$ scales as $D \sim 1/N$. This may
be seen by considering the non-interacting case in which the
magnetization evolves by pure diffusion, where the diffusion
constant scales in this form. Since we are interested in a
thermodynamically unstable $m=0$ state, corresponding to a local
minimum of the entropy, we may, for simplicity, consider an entropy
function of the form \be
 s(m)=am^2-bm^4,
\l{toy_entropy} \ee with $a$ and $b$ non-negative parameters. The
Fokker-Planck equation for the probability distribution $P(m,t)$ at
time $t$ reads \be \frac{\partial P(m,t)}{\partial t} =
D\frac{\partial^2 P(m,t)}{\partial m^2} -\frac{\partial}{\partial
m}\left(\frac{\partial s}{\partial m}P(m,t)\right). \l{FPE} \ee This
equation may be viewed as describing the motion of an overdamped
particle with coordinate $m$ in a potential $-s(m)$ at a temperature
$T=D$. To probe the relaxation process from the $m=0$ state, it is
sufficient to consider the entropy in Eq. (\ref{toy_entropy}) with
$b=0$. With the initial condition, $P(m,0)=\delta(m)$, the long-time
distribution is \cite{Risken} \be P(m,t) \sim \exp \left[
-\frac{ae^{-at}m^2}{D} \right]. \ee It follows that the relaxation
time from the unstable state, $\tau_{us}$, which corresponds to the
width reaching a value of $O(1)$, satisfies \be \tau_{us} \sim -\ln
D \sim \ln N. \ee Similar behavior has been found for the model in
Eq. (\ref{hamex}) with Metropolis-type canonical dynamics at fixed
temperature \cite{Mukamel_2005}. Thus, the logarithmic divergence
with $N$ of the relaxation time seems to be independent of the
nature of the dynamics (i.e., whether microcanonical or canonical).

The relaxation  process from a metastable state (rather than an
unstable state discussed above) has also been studied in the past \cite{Mukamel_2005}. Here the entropy has a local maximum at $m=0$, while the global
maximum occurs at some $m \ne 0$. As one would naively expect, the
relaxation time $\tau_{ms}$ from the metastable $m=0$ state grows
exponentially with $N$: $\tau_{ms} \sim e^{N\Delta s}$
\cite{Mukamel_2005}. The entropy barrier $\Delta s$ corresponding to
the non-magnetic state is the difference in entropy between that of
the $m=0$ state and the entropy at the local minimum separating it
from the stable equilibrium state. Such exponentially long
relaxation times are expected to occur independently of the nature
of the order parameter or of the type of the dynamics (stochastic
or deterministic). This has been found in the past in numerous
studies of canonical, Metropolis-type dynamics, of the Ising model
with mean-field interactions \cite{Griffiths_1966}, in deterministic
dynamics of the $XY$ model \cite{Antoni_2004}, and in models of
gravitational systems \cite{Chavanis_2003,Chavanis_2005}.

\section{Weak long-range interactions}
\l{weak} Here we consider systems with weak long-range
interactions, with $\sigma>0$. These systems are additive and thus,
have usual properties as for systems with short-range interactions.
This is true unless one is close to a phase transition, where, as
mentioned in the Introduction, building up of long-range
correlations leads to modification of the thermodynamic properties.
For example, critical exponents near a continuous transition become
dependent on the interaction parameter $\sigma$. In this Section, we
briefly discuss the upper critical dimension $d_c(\sigma)$ for these
systems above which the critical exponents assume the Landau or
mean-field values. For details, see \cite{Mukamel_2009}.

We first discuss the case of short-range interactions. We start with
the coarse-grained Landau-Ginzburg effective Hamiltonian of the
system. This Hamiltonian involves only the long-wavelength degrees
of freedom, and is obtained by averaging over the short-wavelength
ones. For a given system, one obtains this Hamiltonian
phenomenologically from the symmetry properties of the order
parameter involved in the transition. In systems with a single
component, Ising-like order parameter, say, the magnetization, the
effective Hamiltonian involves the local coarse-grained
magnetization $m({\bf r})$ defined at the spatial location {\bf r}.
If the interaction is short-ranged, this Hamiltonian has the form
\be \beta H= \int d^d{\bf r}\left[\frac{1}{2}tm^2 + \frac{1}{4}um^4
+\frac{1}{2}(\nabla m)^2\right], \l{LGham} \ee where $t$ and $u>0$
are phenomenological parameters, and $d$ is the spatial dimension.
Close to the critical temperature $T_c$, the parameter $t$ may be taken to depend on temperature, $t \propto
\fr{(T-T_c)}{T_c}$.
Considering the model at $t>0$, when the equilibrium state is a
paramagnetic one, we first evaluate the fluctuations of the order
parameter around its average value, $\langle m \rangle=0$.
Expressing the effective Hamiltonian in terms of the Fourier modes
of the order parameter, $m({\bf q})=\int d^d{\bf r} e^{i{\bf q}\cdot
{\bf r}}m({\bf r})$, and neglecting the fourth-order term in Eq.
(\ref{LGham}) close to the transition, the Hamiltonian in the
thermodynamic limit reduces to that of a Gaussian model. Thus, one has
\be \beta H
=\frac{1}{2 (2\pi)^d} \int d^d{\bf q} (t+q^2)m({\bf q})m(-{\bf q}).
\ee From this, it follows that the two-point correlation function,
$\langle m({\bf r}) m({\bf r'})\rangle$, is given by \be \langle
m({\bf r})m({\bf r'}) \rangle =\frac{1}{(2 \pi)^d} \int d^d{\bf q}
\frac{e^{-i{\bf q}\cdot{\bf (r-r')}}}{t+q^2}. \l{mcorr} \ee On
scaling $q$ by $\sqrt t$, the integral in Eq. (\ref{mcorr}) implies
a correlation length $\xi=t^{1/2}$. At distances much larger than
$\xi$, the correlation function decays exponentially as $e^{-{\bf
|r-r'|}/ \xi}$, with a sub-leading power law correction.

From Eq. (\ref{mcorr}), integrating over modes with wavelengths
bigger than the correlation length $\xi$, one gets \be \langle
m^2({\bf r}) \rangle = \int_{q<{\sqrt t}} \frac{d^d{\bf
q}}{(2\pi)^d} \frac{1}{t+q^2} ~ \propto ~ t^{\frac{d}{2}-1} ~.
\label{m^2_t>0} \ee In the Landau or mean-field theory, one neglects
fluctuations of the order parameter. To find the dimensions at which
this assumption is valid, let us first consider the fluctuations of
the order parameter about its average below the critical point
($t<0$) by writing $m({\bf r})$ as $m({\bf r})=m_0 +\delta m({\bf
r})$, where the average $m_{0} = \sqrt{-\frac{t}{u}}$. With this
form for $m({\bf r})$, from Eq. (\ref{LGham}), it follows that, to
second order in $\delta m({\bf r})$, the Landau-Ginzburg effective
Hamiltonian close to the transition point is given by \be \beta H
=\int d^d{\bf r}\left[ \fr{1}{2}|t|(\delta m({\bf r}))^2 +
\frac{1}{2}(\nabla \delta m({\bf r}))^2\right]. \l{LGhamfl} \ee We
now see that the fluctuations of the order parameter around the
average value below but close to the transition obey a Hamiltonian
similar to that for the fluctuations above and close to the
transition. It then follows from Eq. (\ref{m^2_t>0}) that $\langle
\delta m^2({\bf r}) \rangle ~ \propto ~ |t|^{\frac{d}{2}-1}$. For
the Landau theory to be self-consistent near the transition,
fluctuations of the order parameter should be negligibly small
compared with the average value of the order parameter, $\langle
\delta m^2({\bf r}) \rangle \ll m_0^2$, which is true so long as $d$
is greater than the upper critical dimension $d_c=4$. In dimensions
less than $4$, negligible fluctuations of the order parameter around
its average can be achieved only away from the critical point for
$|t| > |t_G|$, where $t_G$ defines the Ginzburg temperature
interval. This is known as the Ginzburg criterion
\cite{Ginzburg_1960}, \cite{Hohenberg_1968},
\cite{Als-Nielsen_1977}.

Extending this analysis to systems with weak LRI, one finds that the
Landau-Ginzburg effective Hamiltonian takes the form \be
\beta H= \int d^d{\bf r}\left[\frac{1}{2}tm^2 + \frac{1}{4}u m^4\right]+ \int d^d{\bf r} ~ d^d{\bf r'} m({\bf r})m({\bf r'}) \frac{1}{|{\bf r-r'}|^{d+\sigma}},
\ee where the second integral accounts for the contribution from
the long-range interaction potential to the energy. Next, we note
that, to leading order in $q$, the Fourier transform of the
long-range potential is of the form $a+bq^\sigma$, where $a$ and $b$
are constants (For integral $\sigma$, a logarithmic correction is
present, e.g.,  $a+bq^2 \ln q$ for $\sigma =2$; however, these
logarithmic corrections do not affect the conclusions reached
below.). In terms of the Fourier components of the order parameter,
the Landau-Ginzburg effective Hamiltonian close to the transition
thus takes the form \be \beta H = \frac{1}{2 (2\pi)^d}\int d^d{\bf
q}(\bar{t}+bq^\sigma+q^2) m({\bf q})m(-{{\bf q}}), \l{LRham} \ee
where $\bar{t}=t+a$.

To obtain the upper critical dimension, one may now perform an
analysis similar to that discussed above for systems with
short-range interactions. For $\sigma > 2$, the $q^\sigma$ term in
Eq. (\ref{LRham}) may be neglected in comparison to the $q^2$ term.
One then recovers the behavior for models with short-range
interactions and the upper critical dimension is $d_c=4$. On the
other hand, for $0<\sigma<2$, the dominant term is $q^\sigma$. The
correlation length in this case diverges as $\xi~ \propto
~|\bar{t}|^{-1/\sigma}$. The order parameter fluctuations satisfy
\be \langle (\delta m({\bf r}))^2 \rangle= \int_{q<1/\xi}
\frac{d^d{\bf q}}{(2 \pi)^d} \frac{1}{|\bar{t}|+q^\sigma} ~ \propto
~ |\bar{t}|^{d/\sigma-1} ~. \ee Requiring that the fluctuations are
much smaller than the average value of the order parameter
$m_0=\sqrt{-\fr{\bar{t}}{u}}$, one concludes that the upper critical
dimension is $d_c=2\sigma$. Thus, in $d<4$ dimensions, there exists
a critical $\sigma$, given by $\sigma_c(d)=\frac {d}{2}$, such that,
for $0< \sigma < \sigma_c(d)$, the critical exponents are mean-field
like.This naive approach suggests that, for $\sigma_c(d) < \sigma <
2$, the critical exponents are affected by the long-range nature of
the interaction and hence, become dependent on $\sigma$, see
\cite{Fisher_1972}. In fact, it has been shown using Renormalization
Group in $d=2\sigma-\epsilon$ dimensions that the value of $\sigma$
above which short-range behavior is recovered is smaller than $2$,
and depends on the dimension $d$ \cite{Sak_1973}.

These conclusions are schematically shown in Fig. \ref{Fig7}, where
the associated critical behavior in different regimes are also
indicated. \bef
\begin{center}
\includegraphics[scale=0.4]{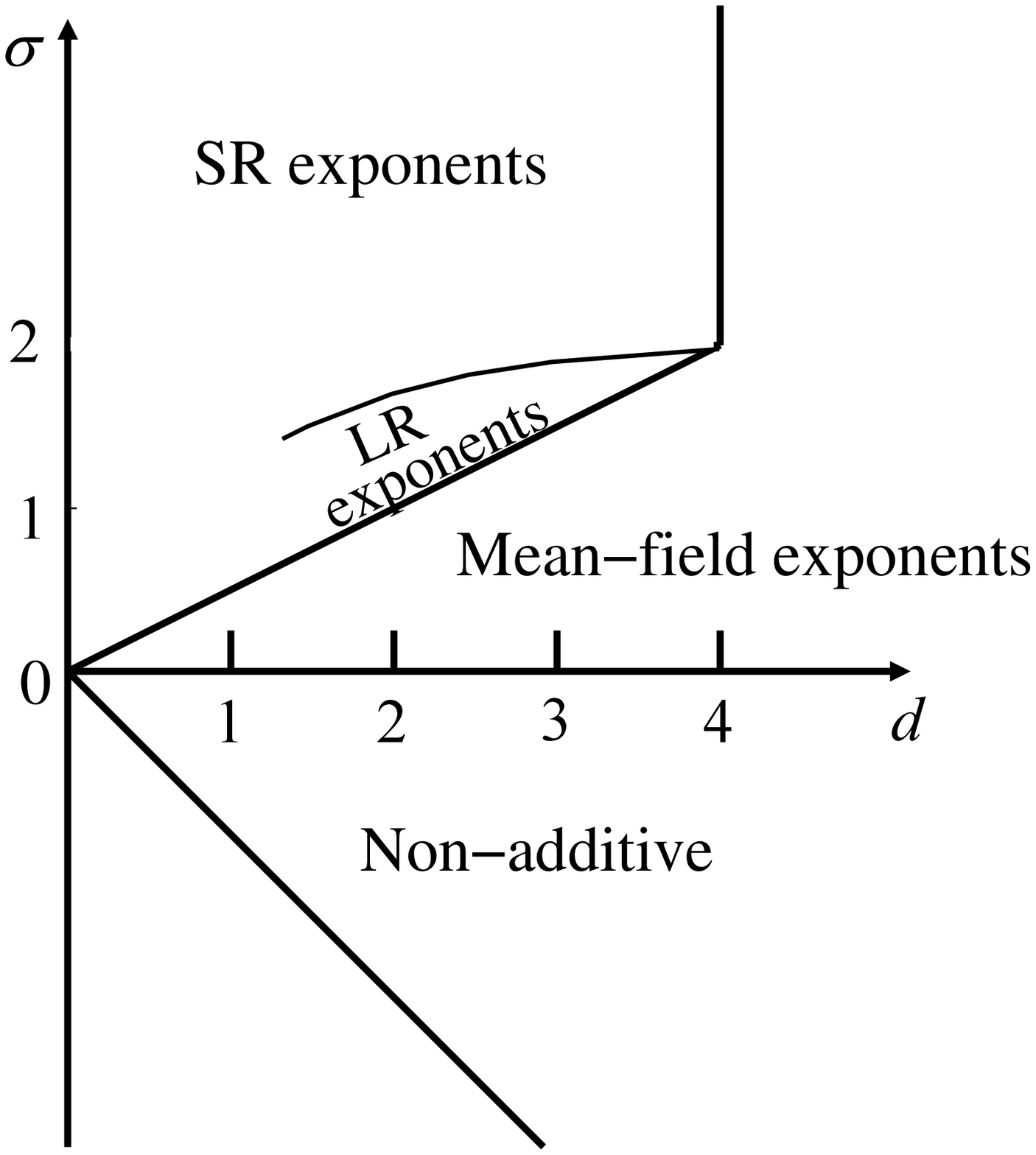}
\caption{The schematic $(d,\sigma)$ phase diagram, showing various regions with different critical behavior. Here LR stands for long-range, while SR stands for short-range. The system is non-additive with strong long-range interactions for $d \le \sigma \le 0$. For $\sigma > 0$, the critical exponents can be either mean-field like, short-range like or characteristic of the long-range interactions, depending on $\sigma$ and $d$. The line separating the LR from the SR behavior is indicated close to the point $d=2\sigma=4$, where it has been evaluated using Renormalization Group calculations in $d=2\sigma-\epsilon$ dimensions. Note that for the case $\sigma>2$, no phase transition takes place for $d \le 1$.}
\l{Fig7}
\end{center}
\eef

\section{Nonequilibrium steady states and long-range interactions}
\l{sec:NESS-Local}

We now turn to discuss steady states in systems driven out of
equilibrium. In these systems, the drive may be provided either by an
external force, such as due to an electric field, or by coupling to external thermostats at different temperatures. We consider here systems for which the dynamics is conserving and local. These systems are typically characterized
by long-range correlations which have been shown to lead to phase transitions and long-range order in many one-dimensional models. For reviews on steady state
properties of driven models, see, for example, \cite{Zia_1995},
\cite{Schutz_2000}, \cite{Mukamel_2001}, \cite{Derrida_2007}. In
some models, features characteristic of strong LRI, such as
inequivalence of ensembles, have been observed
\cite{Grosskinsky_2008}. In this Section, we discuss a particular
model with local dynamics, called the $ABC$ model, which exhibits
spontaneous symmetry breaking in one dimension and for which such
correlations can be explicitly demonstrated
\cite{Evans_1998a,Evans_1998b}. Moreover, for particular values of
the parameters defining this model, the steady state becomes an
equilibrium state obeying detailed balance. The weights of
configurations in such a state are given by an effective Hamiltonian,
which has explicit long-range interacting terms.

The model is defined on a one-dimensional lattice of $N$ sites with
periodic boundary conditions. Each site is occupied by either
an $A$, a $B$, or a $C$ particle. Configurations evolve by random
sequential dynamics as follows: at each time step, two neighboring
sites are chosen randomly and the particles on these sites are exchanged
according to the following rules:
\be
\begin{picture}(130,37)(0,2)
\unitlength=1.0pt \put(36,6){$BC$} \put(56,4) {$\longleftarrow$}
\put(62,-1) {\footnotesize $1$} \put(56,8) {$\longrightarrow$}
\put(62,14) {\footnotesize $q$} \put(80,6){$CB$} \put(36,28){$AB$}
\put(56,26) {$\longleftarrow$} \put(62,21) {\footnotesize $1$}
\put(56,30) {$\longrightarrow$} \put(62,36) {\footnotesize $q$}
\put(80,28){$BA$} \put(36,-16){$CA$} \put(56,-18) {$\longleftarrow$}
\put(62,-23) {\footnotesize $1$} \put(56,-14) {$\longrightarrow$}
\put(62,-8) {\footnotesize $q$} \put(80,-16){$AC$.}
\end{picture}
\label{ABCrules}
\ee
The rates are cyclic in $A$, $B$ and $C$ and conserve the total number
of particles $N_A, N_B$ and $N_C$ of each type, respectively.

For $q=1$, the particles undergo symmetric diffusion. At long times,
the system reaches an equilibrium steady state, which is
disordered. However, for $q \neq 1$, the particle exchange rates
are biased, and the system settles into a nonequilibrium steady state
which shows separation of the particle species into three distinct
domains in the thermodynamic limit.

To be specific, we take $q<1$, although the analysis is easily
extended to any $q \ne 1$. In this case, the bias drives, say,
an $A$ particle to move to the left inside a $B$ domain, and to
the right inside a $C$ domain. Therefore, starting with a random
initial configuration, after a relatively short time, the system
reaches a configuration of the type $\ldots AABBCCAAAB \ldots$ in
which $A,B$ and $C$ domains are located to the right of $C$, $A$
and $B$ domains, respectively.  Due to the bias $q <1$, the domain
walls $\ldots AB \ldots$, $\ldots BC \ldots$, and $\ldots CA \ldots$,
are stable, and configurations of this type are long-lived. In fact,
the domains in these configurations diffuse into each other and coarsen
on a timescale of the order of $q^{-l}$, where $l$ is the typical domain size.
This coarsening process leads to the growth of the typical domain size as
$( \ln t)/\vert\ln q \vert$. Eventually, the system settles into a
phase-separated state of the form $A \ldots AB \ldots BC \ldots C$.
A finite system does not stay in such a state indefinitely.  For example,
the $A$ domain breaks up into smaller domains in a time of order
$q^{-\mathrm{min} \lbrace N_B,N_C \rbrace}$. In the thermodynamic
limit, however, when the density of each type of particles is
non-vanishing, the timescale for the break up of extensive domains
diverges and the system remains in the phase-separated state forever.
Generically, the system supports particle currents in the steady state.
This can be seen by considering, say, the $A$ domain in the phase-separated
state. The rates at which an $A$ particle traverses a $B$ ($C$) domain to
the right (left) is of the order of $q^{N_B}$ ($q^{N_C}$), so that the net
current is of order $q^{N_B}-q^{N_C}$, vanishing exponentially with $N$.
This implies that, for the special case of equal densities of the three
particle species, $N_A=N_B=N_C$, the current is zero for any system size.

One finds that, for the special case of equal densities, $N_A=N_B=N_C$,
the dynamics satisfy {\it detailed balance} with respect to a
distribution function. The model in this case reaches an equilibrium
steady state. It turns out however that, although the dynamical rules
of the model are {\it local}, the effective Hamiltonian corresponding
to this equilibrium steady state has {\it
long-range interactions}, and thus, supports phase separation,
consistent with our predictions above.

In order to specify the probability distribution of configurations
for equal densities, we define a local occupation variable
$\lbrace X_i \rbrace = \lbrace A_i,B_i,C_i \rbrace$, where $A_i$, $B_i$ and $C_i$
are equal to one if site $i$ is occupied by an $A$, a $B$, or a $C$ particle,
respectively, and is zero otherwise. The probability of finding the system in a
configuration $\lbrace X_i \rbrace$ is given by
\be
W_N(\{X_i\}) = Z_N^{-1}q^{{H}(\lbrace X_i\rbrace)},
\label{ABCweight}
\ee
where the effective Hamiltonian $H$ is given by
\be
H(\{ X_i \})= \sum_{i=1}^{N-1}\sum_{k=i+1}^{N} (C_i B_{k} + A_i C_{k} + B_i A_{k})  - (N/3)^2,
\l{ABCham}
\ee
and the normalization or the partition function $Z_N$ is given by
$ Z_N=\sum q^{{H} (\lbrace X_i \rbrace)}$. In this Hamiltonian,
the site $i=1$ is arbitrary and can be chosen as any other site on
the ring, since the Hamiltonian does not depend on this choice. The
Hamiltonian involves strong long-range interactions, where the strength
of the interaction between two sites is independent of the separation
(thus, $\sigma=-1$). Also, the Hamiltonian is non-extensive, with
energy scaling as $N^2$. It may be verified that the dynamics
(\ref{ABCrules}) satisfy detailed balance with respect to the probability
distribution in Eq. (\ref{ABCweight}), with the Hamiltonian in
Eq. (\ref{ABCham}) \cite{Mukamel_2009}.

The $ABC$ model exhibits phase separation and long-range order so
long as $q\ne 1$. The parameter $q$ acts like the temperature for
the case of equal densities, with $\beta=-\ln q$, as can be seen
from Eq. (\ref{ABCweight}). A very interesting limit is that of
infinite temperature, with $q \rightarrow 1$. To probe this limit,
Clincy and co-workers \cite{Clincy_2003} studied the case $q=e^{-\beta/N}$,
which amounts to either scaling the temperature by $N$, or, alternatively,
scaling the Hamiltonian in Eq. (\ref{ABCham}) by $1/N$, as is done in the prescription due to Kac (Section \ref{thermodynamics}). In this case, the model shows a phase transition from
a homogeneous phase at high temperatures to a phase-separated one at
low temperatures across the critical point $\beta_c=2 \pi \sqrt{3}$.

In this section, we discussed how, in the $ABC$ model for equal particle
densities, the effective Hamiltonian governing the steady state involves
explicitly strong long-range (in fact, mean-field) interacting terms. By
continuity, this is expected
to hold even for the case of non-equal densities, although, in such cases,
no effective Hamiltonian could be explicitly written. It remains to explore
in more detail steady state properties of driven systems within the framework
of systems with long-range interactions.

\section{Conclusions}
In this paper, we reviewed the thermodynamic and dynamic properties of systems with long-range
pairwise interactions (LPI) decaying as $1/r^{d+\sigma}$ at large
distances $r$ in $d$ dimensions. Systems with a slow decay of the
interactions, termed ``strong'' LRI, have superextensive energy. These systems are characterized by unusual properties such as inequivalence of ensembles, negative specific
heat, slow decay of correlations, anomalous diffusion and ergodicity breaking. Systems with faster decay of the interaction potential, termed weak ``LRI'', have
additive energy, thus resulting in less dramatic effects. These interactions affect the thermodynamic behavior of systems near phase transitions, where long-range correlations are naturally present. We also discussed long-range correlations in systems driven out of equilibrium when the dynamics involves conserved quantities.

\section*{Acknowledgments}

This work was supported by the Minerva Foundation with funding from
the Federal German Ministry for Education and Research, and by the
French \textit{Agence Nationale de la Recherche} through the ANR
program STATFLOW (ANR-06-JCJC-0037-01) and through the ANR program
STATOCEAN (ANR-09-SYSC-014).

\bibliographystyle{elsart-num}
\bibliography{Mukamel,FBouchet,FBouchet-Proceedings,FBouchet-Books,Long_Range,Meca_Stat_Euler,Experimental_2D_Flows,Euler_Stability,Stochastic_Processes}

\end{document}